\documentclass{PoS}

\usepackage{lineno} 
\usepackage{atlasphysics}
\usepackage[labelformat=simple]{subcaption}

\title{Measurements of the production of single top quarks or of top quark pairs in association with a W~boson, a Z boson, or a photon with ATLAS \& CMS}

\ShortTitle{Production of Single Top Quarks or of Top Quark Pairs in Association with $\Wboson$,$\Zboson$, or $\gamma$}

\author{\speaker{T. G. McCarthy}, on behalf of the ATLAS and CMS Collaborations \\
        	Max-Planck-Institut f$\ddot{\textrm{u}}$r Physik, F$\ddot{\textrm{o}}$hringer Ring 6, 80805 M$\ddot{\textrm{u}}$nchen, Germany\\
        E-mail: \email{tmccarth@mpp.mpg.de}}

\abstract{These proceedings highlight a selection of recent ATLAS and CMS measurements which target either the production of a single top quark or the associated production of a top-antitop quark pair together with a $\Wboson$ boson, a $\Zboson$ boson, or a photon ($\gamma$). The highlighted analyses are based predominantly on proton-proton collision data collected by the two experiments during Run 2 of the LHC at a centre-of-mass energy of $\sqrt{s}=13$ \TeV \ with datasets corresponding to integrated luminosities of up \mbox{to 77.5 \ifb}. One combination of ATLAS and CMS single top quark measurements employing the full Run 1 dataset based on $\rts=7,8 \ \TeV$ is also included. Such measurements probe the Standard Model couplings of top quarks to gauge bosons and offer sensitivity to a number of anomalous non-Standard-Model effects.}

\FullConference{7th Annual Conference on Large Hadron Collider Physics - LHCP2019\\
		20-25 May, 2019\\
		Puebla, Mexico}

\begin{document}
\section{Introduction}

A rich and diverse set of measurements have been performed by both the ATLAS~\cite{ATLASExperiment} and CMS~\cite{CMSExperiment} Collaborations which probe the Standard Model (SM) predictions associated with top quark production and decay at the Large Hadron Collider (LHC).  In the following proceedings a particular focus is placed on measurements which probe the coupling of top quarks to some of the SM bosons, specifically in the context of the production of top-antitop quark ($t\bar{t}$) pairs together with a $\Wboson$ boson, a $\Zboson$ boson, or a photon ($\gamma$).  Measurements targeting single top or antitop quark production at a centre-of-mass energy of $\rts=13 \ \TeV$ via two of the dominant single-top quark production modes at the LHC ($t$-channel and associated $t+W$ channel) are also discussed, as well as a full ATLAS and CMS combination based on earlier $\rts=7,8 \ \TeV$ data from Run 1 of the LHC.  In addition two recent measurements are presented -- one from ATLAS and one from CMS -- which target $\ttbar$ and $t\Wboson$ production in order to improve the modelling used in simulated datasets, and also in order to look for hints of new physics.  The measurements presented target a variety of final-state signatures based on the decays of the top quarks, or the $\Wboson$ or $\Zboson$ bosons.


\section{Cross-section measurements in $t\bar{t}+ X$ final states ($X = \Wboson,\Zboson,\gamma$)}

\subsection{Associated $\ttbar+\gamma$ production}

\begin{figure}[!htbp]
\begin{subfigure}{0.49\textwidth}
\begin{centering}
\includegraphics[scale=0.36]{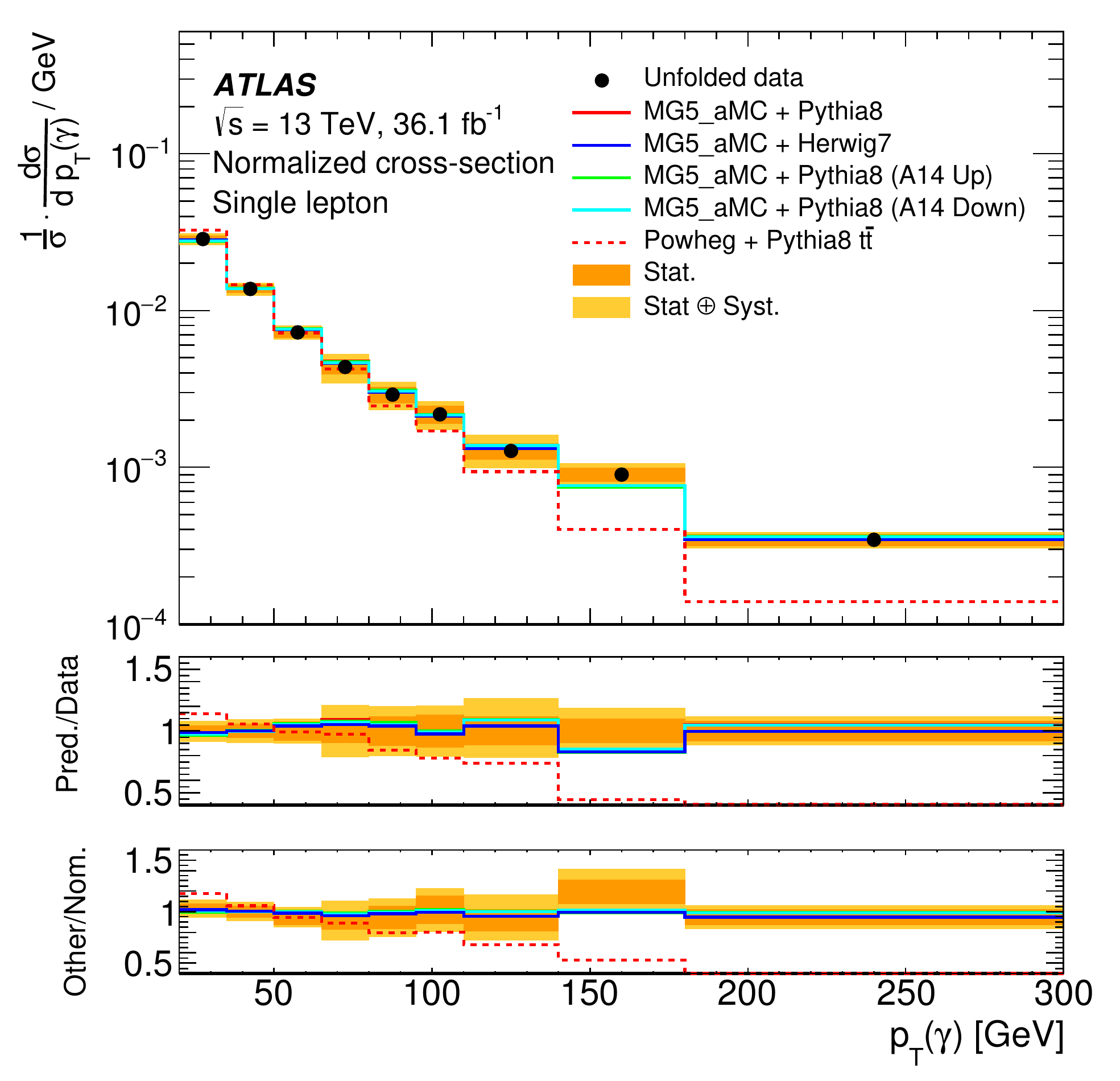}
\caption{}
\label{Fig-ATLAS-tty-a}
\end{centering}
\end{subfigure} 
\begin{subfigure}{0.49\textwidth}
\begin{centering}
\includegraphics[scale=0.36]{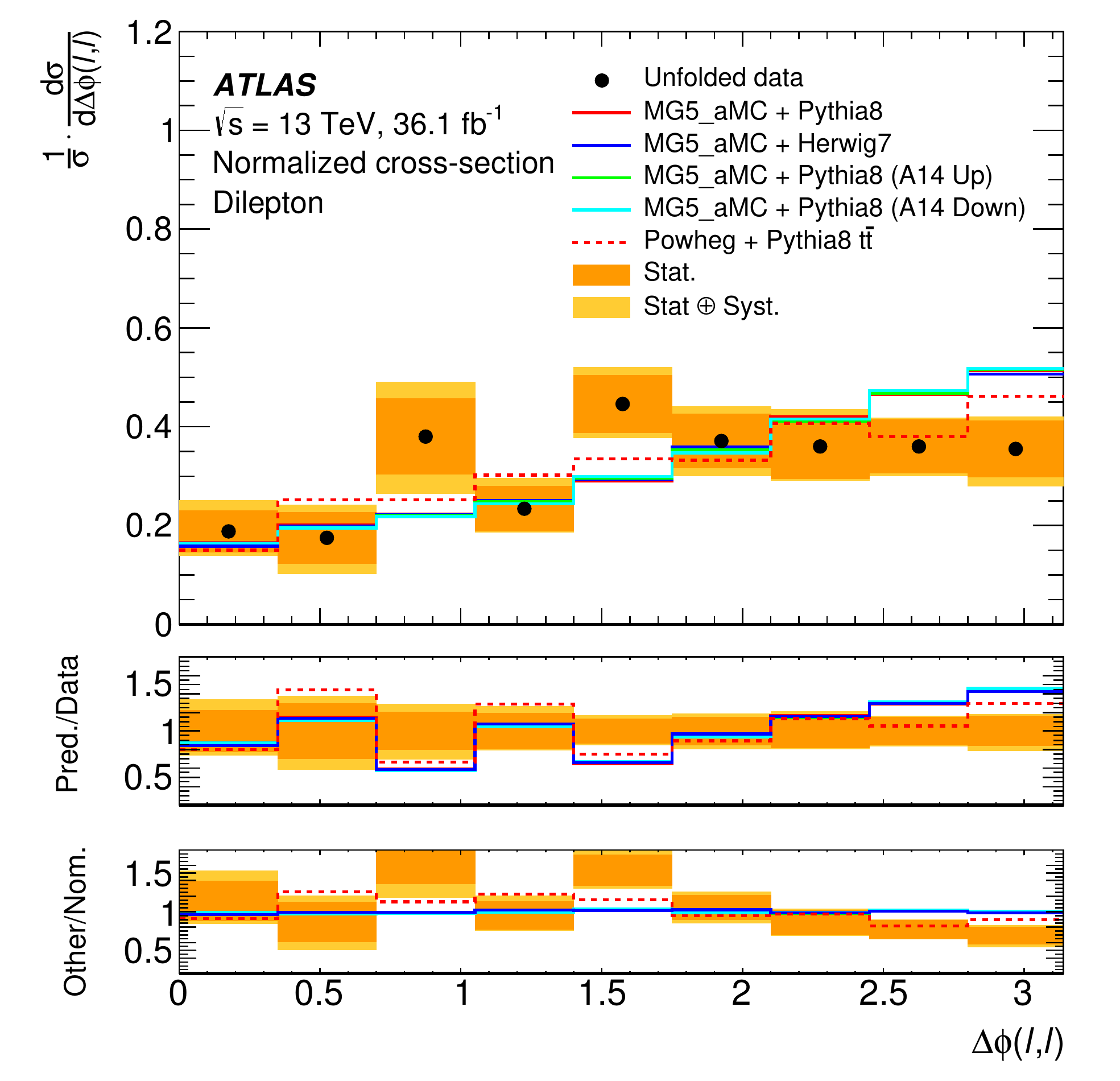}
\caption{}
\label{Fig-ATLAS-tty-b}
\end{centering}
\end{subfigure} 
\caption{Two representative differential cross-sections results from the ATLAS fiducial measurement of $\ttbar\gamma$ production~\cite{ATLAS_tty}.  The normalized results are shown as a function of (a) the photon \pt \ in the single-lepton channel, and (b) the absolute azimuthal angle between the two leptons from the $\ttbar$ system in the dileptonic channel.  Two sets of ratios are displayed bin-by-bin in each of the figures: the central portion shows the ratio of the various predictions to the unfolded results from the measured data, and the lowermost portion shows the ratio of the various predictions to those from the nominal \MGaMCatNLO \ sample.}
\label{Fig-ATLAS-tty-diff}
\end{figure}

\begin{figure}[!htbp]
\begin{centering}
\includegraphics[scale=0.55]{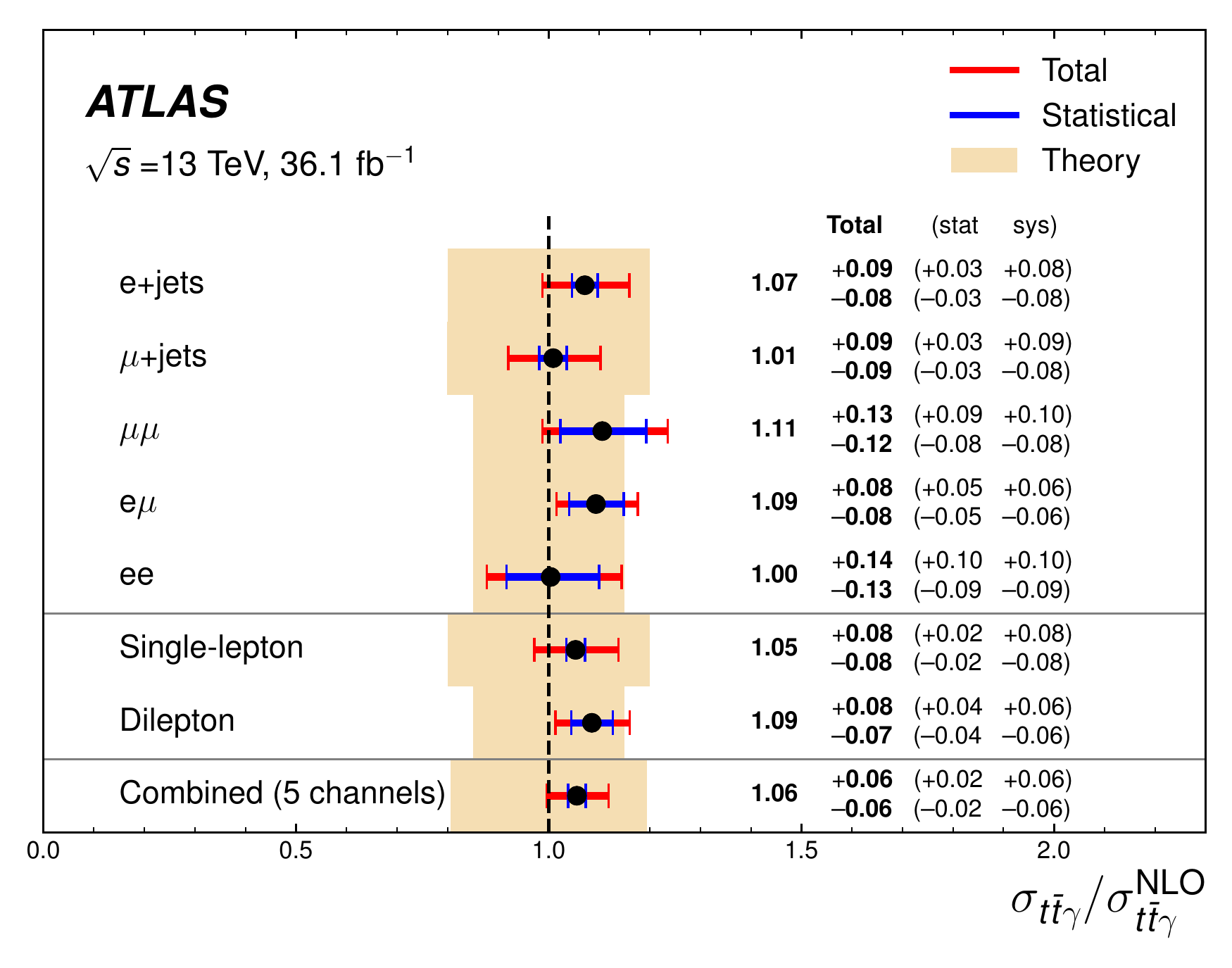}
\caption{A summary of the measured fiducial cross-section results for $\ttbar\gamma$ production based on ATLAS $\rts = 13$ \TeV \ data~\cite{ATLAS_tty}.  Shown are the individual measurements, the combined results inclusive in lepton flavour, and the overall combination based on both the $\ell+$jets and dilepton channels.  All results shown, as well as the combinations, correspond to the same single ATLAS analysis.}
\label{Fig-ATLAS-tty-summary}
\end{centering}
\end{figure}

A recent ATLAS result presents a set of inclusive and differential cross-section measurements for the production of a $\ttbar$ pair together with an isolated photon ($\ttbar\gamma$) \cite{ATLAS_tty}.  The measurements are performed in a fiducial phase-space region\footnote{The fiducial region refers to a set of criteria placed on the photon, such as its transverse momentum, pseudorapidity, and isolation in relation to other objects based on Monte Carlo truth information. For complete details refer to \cite{ATLAS_tty}.} and target both semileptonic and dileptonic $\ttbar$ final states\footnote{According to the SM, each of the (anti)top quarks decay in nearly 100\% of cases via $t\to\Wboson b$.  The semileptonic and dileptonic final states refer to the subsequent decays of the two $\Wboson$ bosons; in the dileptonic final state both $\Wboson$ bosons decay leptonically via $W\to\ell\nu_{\ell}$, whereas in the semileptonic final state one $\Wboson$ boson decays leptonically, and the other hadronically via $\Wboson\to\qqbar$.}.  Dedicated neural networks\footnote{The neural-network algorithms employed in the analysis are feedforward binary classifiers.  Additional detail and further references can be found in \cite{ATLAS_tty}.} are trained separately for the two targeted $t\bar{t}$ decay channels and are employed to suppress the various SM backgrounds.  The inclusive fiducial cross sections in the semileptonic (SL) and dileptonic (DL) channels were found to be $\sigma_{\rm fid.}^{\rm SL} = 521 \ \pm \ 9 \ {\rm (stat.)} \ \pm 41 \ {\rm (syst.)}$ fb and $\sigma_{\rm fid.}^{\rm DL} = 69 \ \pm \ 3 \ {\rm (stat.)} \ \pm 4 \ {\rm (syst.)}$ fb, respectively, both in agreement with next-to-leading order (NLO) quantum chromodynamics (QCD) predictions \cite{ATLAS_tty_NLO_predictions}.  The precision in the dileptonic channel is limited by the available data statistics, whereas in the semileptonic channel the largest uncertainties are associated with the measurement of jets and the background modelling.  The differential cross-section measurements were unfolded in order correct for detector-level effects and acceptance losses.  Two selected normalized cross-section results are shown: as a function of the photon transverse momentum (\pt) in the $\ell +$ jets channel\footnote{$\ell+$ jets refers to the $\ttbar$ semileptonic decay channel in which one (anti)top quark decays via $t\to b\qqbar$ and the other via $t\to b\ell\nu_{\ell}$.  The term jets, related to the hadronic decay $\Wboson\to\qqbar$, corresponds to the fact that a quark or antiquark will typically lead to a reconstructed jet. In some later instances in these proceedings the particular lepton flavour targeted is named explicitly, for example when the targeted lepton is an electron ($e+$ jets), or a muon ($\mu+$ jets).} (Figure~\ref{Fig-ATLAS-tty-a}), and as a function of the absolute angular separation in the transverse plane between the two leptons from the $\ttbar$ decay in the dilepton channel (Figure~\ref{Fig-ATLAS-tty-b}).  The latter features the largest disagreement between data and the various configurations of the leading-order (LO) \MGaMCatNLO \ v2.33 \cite{ATLAS_aMCatNlo} predictions, which were otherwise seen to describe the data well. The NLO $t\bar{t}$ prediction based on \PowhegBox \ v2 \cite{ATLAS_PowhegBox} interfaced with \Pythia \ v8.210 \cite{ATLAS_Pythia8} can be seen to more poorly model the $\ttbar\gamma$ signal, particularly at low and high photon \pT.  Figure~\ref{Fig-ATLAS-tty-summary} shows a summary of the measured signal strengths (the ratio of the measured production cross section to the theoretically predicted value, $\mu = \sigma_{t\bar{t}\gamma}/\sigma^{\rm NLO}_{t\bar{t}\gamma}$) in flavour-specific channels based on the targeted leptons.  Also shown are the results from the channel-specific combinations as well as an overall combination of the semileptonic (labelled as ``Single-lepton") and dileptonic results.  The shaded bands correspond to the total uncertainty on the theoretically predicted values, where the largest uncertainties are associated with the choice of QCD scale.

The observation of the rarer single-top quark equivalent process ($t\gamma$) using $\rts=13 \ \TeV$ data and by targeting events featuring a single reconstructed muon together with hadronic jets \cite{CMS_ty} was also recently made by the CMS Collaboration.  The result was featured in another contribution for these proceedings and is therefore not included here.  


\subsection{Associated $\ttbar+\Wboson$ or $\ttbar+\Zboson$ production}

The production rates and kinematics of $\ttbar$ pairs produced in association with heavy electroweak gauge bosons are important to understand given that such processes constitute a significant background for a number of LHC searches featuring high lepton and $b$-jet multiplicities.

The inclusive production cross sections of $\ttbar$ pairs with a $\Wboson$ or a $\Zboson$ boson ($\sigma_{\ttbar\Wboson}$ and $\sigma_{\ttbar\Zboson}$) were measured by the ATLAS Collaboration based on a 36.1 \ifb \ integrated luminosity dataset at $\rts=13 \ \TeV$ \cite{ATLAS_ttV}, building on the earlier result based on 3.2 \ifb \ of integrated luminosity at the same centre-of-mass energy~\cite{ATLAS_ttV_3p2}.  The measurement targeted final states with either two, three or four leptons\footnote{A total of five distinct final states were targeted: 2 opposite-sign leptons ($t\bar{t}\to b\qqbar \bar{b}\qqbar,\ \Zboson\to\ell^+\ell^-$), 2 same-sign leptons ($t\bar{t}\to \ell^{\pm}+$ jets, $\Wboson\to\ell^{\pm}\nu_{\ell}$), 3 leptons ($t\bar{t}\to \ell^{\pm}+$ jets, $\Zboson\to\ell^+\ell^-$ or $t\bar{t}\to b\ell^+\nu_{\ell}\bar{b}\ell^-\nu_{\ell},\ \Wboson\to\ell^{\pm}\nu_{\ell}$), or 4 leptons ($t\bar{t}\to b\ell^+\nu_{\ell}\bar{b}\ell^-\nu_{\ell},\ \Zboson\to\ell^+\ell^-$).} and incorporated a number of orthogonal selections of events in order to exploit signal- or background-dominated regions of phase space.  A representative distribution of the reconstructed $\Zboson$ boson transverse momentum in the most sensitive 3-lepton decay channel of the $t\bar{t}\Zboson$ signal process is shown in Figure~\ref{Fig-ATLAS-ttV-a}.  The predicted signal $t\bar{t}\Zboson$ contribution from a simulated Monte Carlo (MC) dataset generated at NLO with \MGaMCatNLO \ \cite{ATLAS_aMCatNlo} is depicted in light blue in the stacked distribution and the combined SM prediction shows good agreement with the measured data points.  Moreover the figure highlights that already with 36.1 \ifb \ of integrated luminosity one is able to achieve a signal-dominated selection.  Shown in Figure~\ref{Fig-ATLAS-ttV-b} is a two-dimensional contour plot comparing the fitted values of the signal strengths for $\ttbar\Wboson$ and $\ttbar\Zboson$.  The two signal strengths were extracted simultaneously from a profile-likelihood fit, and can be seen to be slightly larger than, but consistent with, an NLO SM prediction~\cite{ATLAS_ttV_theory2}.  The largest sources of uncertainty in the measurement are those associated with the modelling of the signal ($t\bar{t}\Zboson$ and $t\bar{t}\Wboson$) and various background processes.


A first differential $\ttbar\Zboson$ cross-section measurement was performed by the CMS Collaboration with 77.5 \ifb \ of $\rts=13 \ \TeV$ data \cite{CMS_ttZ_diff}.  The measurement focused exclusively on the most sensitive three- and four-lepton final states\footnote{The three- and four-lepton final states correspond to $t\bar{t}\to \ell^{\pm}+$ jets, $\Zboson\to\ell^+\ell^-$ and $t\bar{t}\to b\ell^+\nu_{\ell}\bar{b}\ell^-\nu_{\ell},\ \Zboson\to\ell^+\ell^-$.} and similarly incorporated a number of dedicated signal and control regions which are summarized in Figure~\ref{Fig-CMS-ttZ-SR-CR}.  Both absolute and normalized results, unfolded to correct for detector effects, were presented for two variables sensitive to the $\ttbar\Zboson$ coupling: the transverse momentum of the $\Zboson$ boson, and $\cos\theta^{\star}$, where $\theta^{\star}$ is the angle between the $\Zboson$ boson and the negatively charged lepton from the $\Zboson$ decay in the $\Zboson$ rest frame.  The measured normalized differential cross sections for these two variables are shown in Figure~\ref{Fig-CMS-ttZ-a} and ~\ref{Fig-CMS-ttZ-b}, respectively, where the results are seen to agree with the NLO \MGaMCatNLO \ prediction~\cite{ATLAS_aMCatNlo}.  An extensive set of stringent new limits were able to be placed on the contributions from anomalous non-SM couplings of the top quark to the $\Zboson$ boson, including a set of Wilson coefficients associated with non-SM dimension-six operators in the context of SM effective field theory (EFT) \cite{CMS_ttZ_diff_EFT1,CMS_ttZ_diff_EFT2,CMS_ttZ_diff_EFT3,CMS_ttZ_diff_EFT4}.  In addition to the first set of differential measurements the result yielded the most precise total cross-section measurement of $\sigma_{\ttbar\Zboson} = 1.00 \ {}^{+0.06}_{-0.05} \ {\rm (stat.)} \ {}^{+0.07}_{-0.06} \ {\rm (syst.)}$ pb.  Uncertainties related to the lepton identification efficiency have the greatest impact on the precision of the measurement.  The relative experimental precision of $\frac{\delta\sigma_{\ttbar\Zboson}}{\sigma_{\ttbar\Zboson}} \approx {}^{+9.2\%}_{-7.8\%}$ exceeds that of a recent theoretical NLO prediction \cite{ATLAS_ttV_theory2}. The result complements an earlier CMS inclusive cross-section measurement for both $\ttbar\Wboson$ and $\ttbar\Zboson$ production based on \mbox{35.9 \ifb} of integrated luminosity at the same $\rts = 13 \ \TeV$ centre-of-mass energy \cite{CMS_ttV}.

\vspace{-0.3 cm}

\begin{figure}[!htbp]
\begin{subfigure}{0.47\textwidth}
\begin{centering}
\includegraphics[scale=0.40]{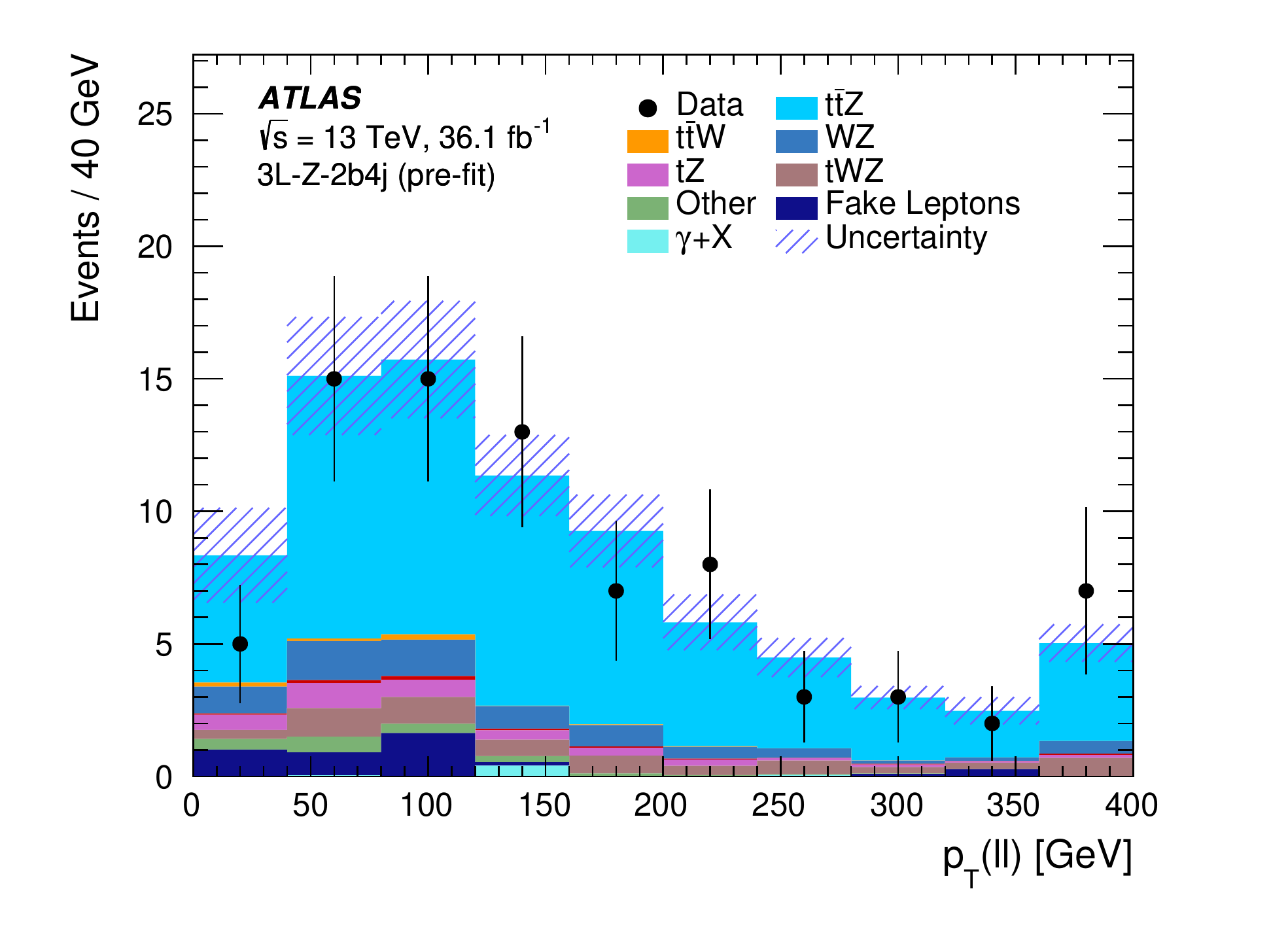}
\caption{}
\label{Fig-ATLAS-ttV-a}
\end{centering}
\end{subfigure} 
\hspace{0.5cm}
\begin{subfigure}{0.47\textwidth}
\begin{centering}
\includegraphics[scale=0.32]{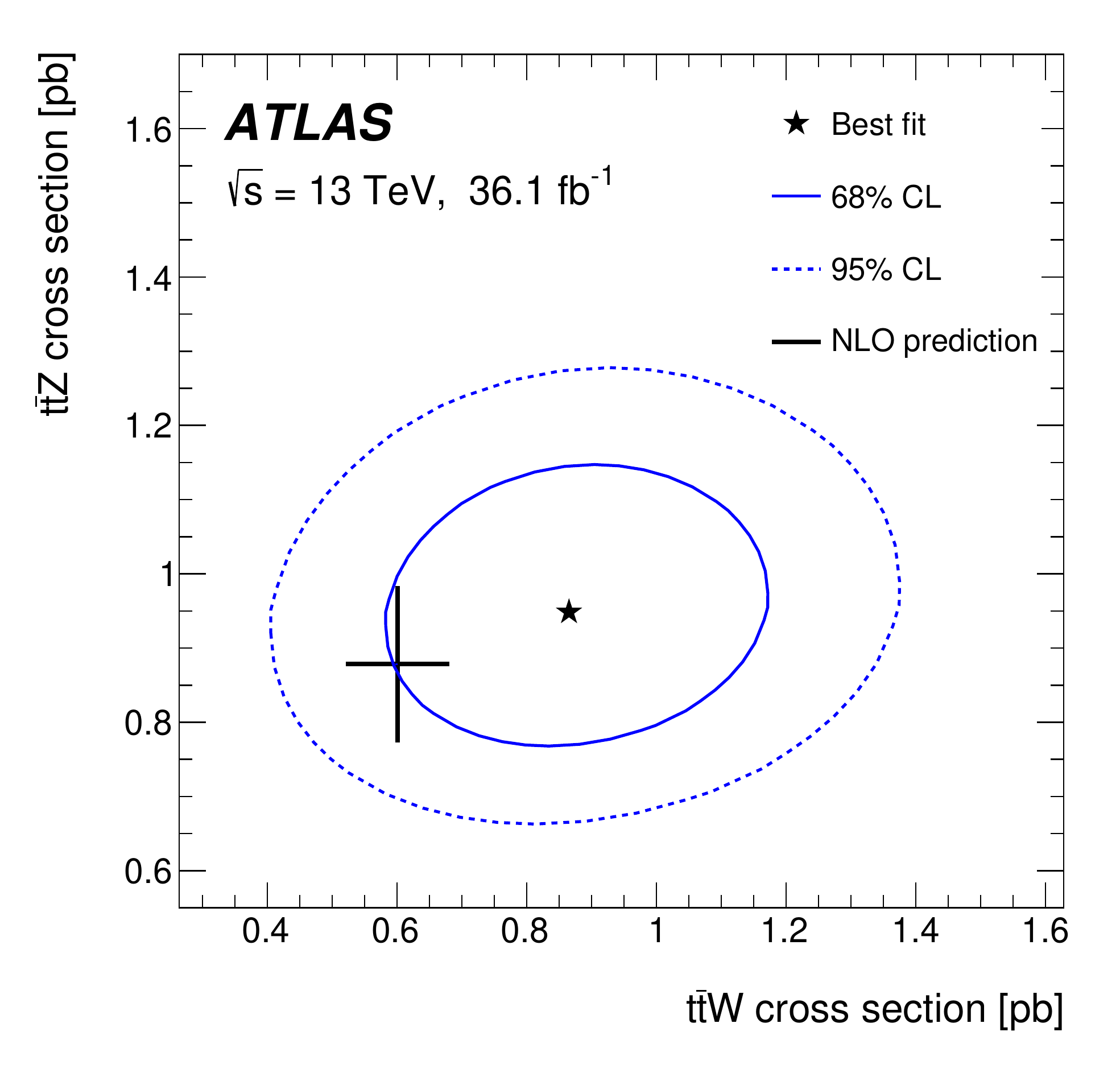}
\caption{}
\label{Fig-ATLAS-ttV-b}
\end{centering}
\end{subfigure} 
\caption{Two figures from the inclusive $\ttbar\Wboson$ and $\ttbar\Zboson$ production cross-section measurement at $\rts = 13$ \TeV \ from ATLAS which targeted several multi-lepton final states based on the decays of the $t\bar{t}$ system and the $\Wboson$ or $\Zboson$ boson, respectively~\cite{ATLAS_ttV}.  Shown are (a) a comparison of the distributions of the measured data and SM prediction for the reconstructed $\Zboson$ boson \pt \ in the most sensitive 3-lepton channel ($t\bar{t}\rightarrow \ell+$ jets, $\Zboson\to \ell\ell$), and (b) the final 2D plot comparing the measured values of the $\ttbar\Wboson$ and $\ttbar\Zboson$ signal strengths from a simultaneous fit to the NLO SM prediction.  The 68\% and 95\% confidence-level (CL) contours are also shown.  The error bars for the NLO prediction include renormalization and factorization scale uncertainties and those associated with variations of the strong coupling constant ($\alphas$).}
\label{Fig-ATLAS-ttV}
\end{figure}

\begin{figure}[!htbp]
\begin{centering}
\includegraphics[scale=0.56]{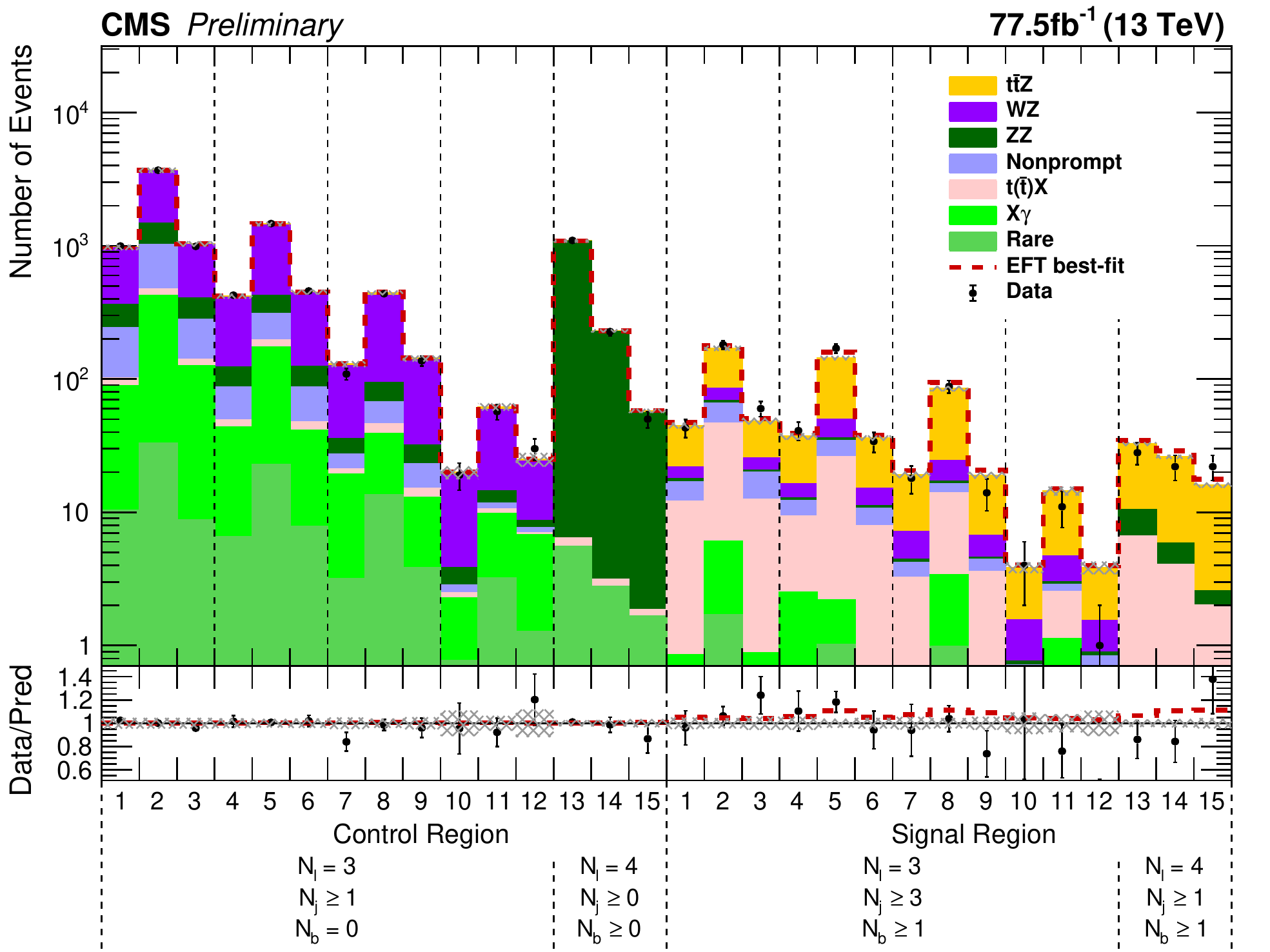}
\caption{A summary of the various signal and control regions used in the differential cross-section measurement for $\ttbar\Zboson$ production performed by the CMS Collaboration~\cite{CMS_ttZ_diff}.  Shown are the total yields for both the measured data and the stacked predictions based on simulations of the various SM processes.  The lower panel shows the ratio of the measured data to the total predicted yields for each bin.  The dashed vertical lines are meant to subdivide groupings of dedicated signal and control  regions into three separate bins of the $\cos\theta^{\star}$ observable described in the text.}
\label{Fig-CMS-ttZ-SR-CR}
\end{centering}
\end{figure}

\begin{figure}[!htbp]
\begin{subfigure}{0.47\textwidth}
\begin{centering}
\includegraphics[scale=0.90]{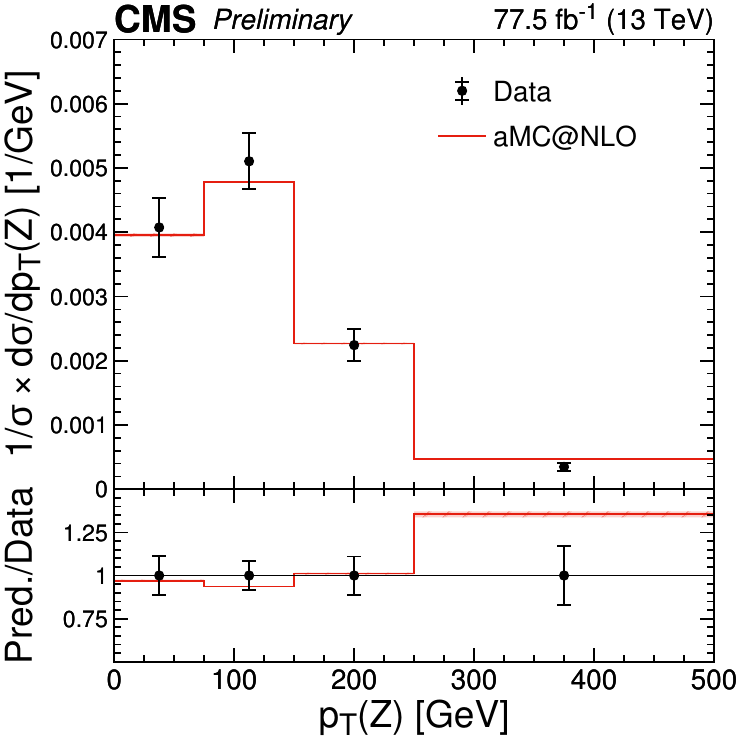}
\caption{}
\label{Fig-CMS-ttZ-a}
\end{centering}
\end{subfigure} 
\hspace{0.5cm}
\begin{subfigure}{0.47\textwidth}
\begin{centering}
\includegraphics[scale=0.90]{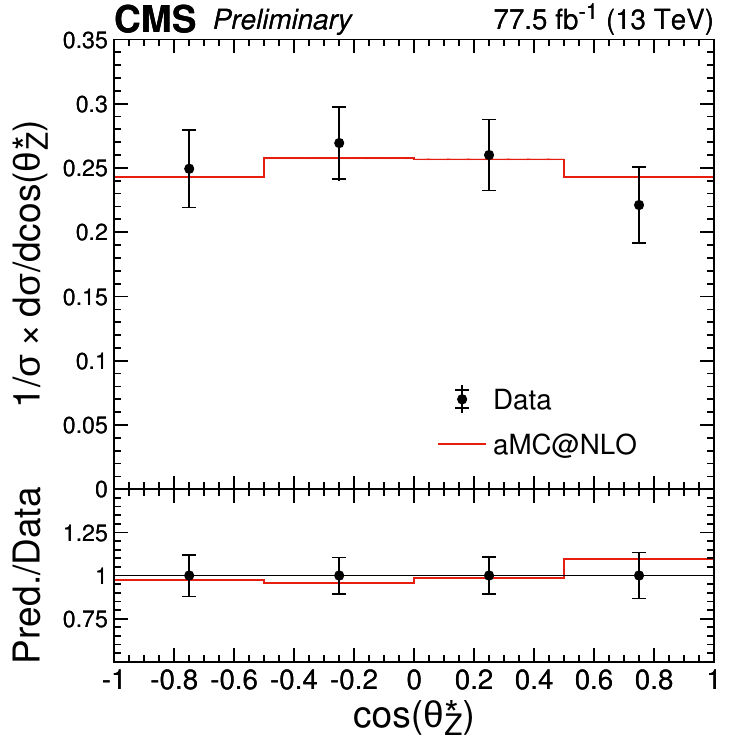}
\caption{}
\label{Fig-CMS-ttZ-b}
\end{centering}
\end{subfigure} 
\caption{Normalized results from the CMS differential cross-section measurement for $\ttbar\Zboson$ production at $\rts=13 \ \TeV$~\cite{CMS_ttZ_diff}.  The measurements are based on 77.5 \ifb \ of LHC data and are presented as a function of (a) the $\Zboson$ boson \pt, and (b) $\cos\theta^{\star}$ as described in the text.  The lower portions of the figures display ratios of the values from prediction relative to those from the unfolded data.}
\label{Fig-CMS-ttZ}
\end{figure}

\clearpage
\section{Measurements of single-top-quark production cross sections}

A number of measurements have been performed by both ATLAS and CMS related to the production of single top or antitop quarks.  The focus for these proceedings was placed on the recent $\rts=13 \ \TeV$ measurements which have been performed for the more dominant leading-order production modes at the LHC, namely via a $t$-channel or via an associated $t+\Wboson$ production diagram.  A recent combination of the Run 1 ($\rts = 7,8 \ \TeV$) measurements from both experiments, including the $s$-channel production mode, which is less dominant at the LHC due to the partonic structure of the colliding protons, is also highlighted.

\subsection{Associated $t+\Wboson$ production}

An inclusive $t\Wboson$ cross-section measurement was performed by CMS with 35.9 \ifb \ of $\rts=13 \ \TeV$ by targeting events with opposite-sign and -flavour two-lepton final states ($e^{\pm}\mu^{\mp}$) together with the presence of at least one $b$-jet\footnote{The $e^{\pm}\mu^{\mp}$ final state corresponds to the $t\Wboson$ decay via $t\Wboson\to \Wboson^{\pm}b\Wboson^{\mp}$, and where each $\Wboson$ boson decays leptonically via $\Wboson\to\ell\nu_{\ell}$. The opposite-flavour lepton selection ($e^\pm \mu^\mp$) reduces the potential for contributions from backgrounds featuring leptonically decaying SM $\Zboson$ bosons.} \cite{CMS_tW}.  Three separate signal regions were constructed, each featuring a low jet multiplicity in order to suppress the contributions from the dominant SM $\ttbar$ background which features a higher average jet multiplicity.  Boosted decision trees (BDT) were used to enhance the signal-to-background ratio.  The BDT output distribution, comparing the measured data with the stacked SM predictions, is shown for the most sensitive signal region in Figure~\ref{Fig-Wt-a} which required exactly one $b$-jet.  In the figure one can note the larger relative contribution of the $t\Wboson$ signal process in the uppermost bins.  A final fitted parameter, translated to a total cross-section value, resulted in a measured value of $\sigma_{t\Wboson} = 63.1 \ \pm \ 1.8 \ {\rm (stat.)} \ \pm \ 6.0 \ {\rm (syst.)} \ \pm \ 2.1 \ {\rm (lumi.)}$ pb, in agreement with an approximate NNLO SM prediction~\cite{CMS_tW_NNLO_prediction}.

Differential cross-section measurements for the same $t\Wboson$ process were measured by the ATLAS Collaboration with 36.1 \ifb \ of integrated luminosity at $\rts=13 \ \TeV$ and based also on opposite-sign two-lepton final states ($\ell^{\pm}\ell^{\mp}$, where $\ell=e,\mu$) \cite{ATLAS_tW}.  The two targeted leptons in this case originate from the leptonic decays of the two $\Wboson$ bosons (decaying via $\Wboson\to\ell\nu_{\ell}$) from the signal process -- one of which being the intermediate $\Wboson$ boson from the (anti)top quark decay.  As in the CMS measurement a dedicated BDT employing a number of discriminating kinematic variables was used to suppress the contributions from the various backgrounds.  The \pt \ of the reconstructed $t\Wboson$ system was seen to offer the highest individual separation power\footnote{For the reconstruction of the $t\Wboson$ system in the transverse plane the missing transverse momentum ($\met$) was used as a proxy to the vector \pt \ sum from the two non-interacting final-state neutrinos.}.  Differential cross-section measurements were produced as a function of a number of variables to probe the kinematics of the $t\Wboson$ system.  The normalized cross-section measurements were unfolded to particle level within a defined fiducial volume.  In general the data were found to exhibit slightly higher-momentum final-state objects compared with the predictions from simulation, as can be seen from the slope in the ratio plot in Figure~\ref{Fig-Wt-b} for the $E(\ell\ell b)$ variable, which corresponds to the energy of the system formed by a 4-vector sum of the two leptons and the bottom quark or antiquark from the top quark decay.  The central ratio plot in the figure compares the measured data with predictions based on two separate simulations with different ways of treating the interference from the associated signal $t\Wboson$ and background $t\bar{t}$ production diagrams: the diagram removal (solid blue line) and diagram subtraction (dashed red line) schemes\footnote{Since $t\bar{t}$ and $t\Wboson$ processes are typically simulated independently, a conventional approach to treat the interference between them is to compare the results based on the predictions for $t\Wboson$ using separate samples which employ so-called diagram removal (DR) and diagram subtraction (DS) techniques and assign this as a systematic uncertainty.  Further information on the diagram removal and subtraction schemes can be found in Ref.~\cite{ATLAS_tW_DR_DS}.}.  In the lowermost ratio plot the predictions from two separate MC generators, \MGaMCatNLO \ v2.2.2 \cite{ATLAS_aMCatNlo} and \PowhegBox \ v1, are compared with the data.  In both cases the MC generators are interfaced with \Herwigpp \ 2.7.1~\cite{ATLAS_tW_Herwigpp} to model the parton shower and hadronization.


\begin{figure}[!htbp]
\begin{subfigure}{0.47\textwidth}
\begin{centering}
\includegraphics[scale=0.35]{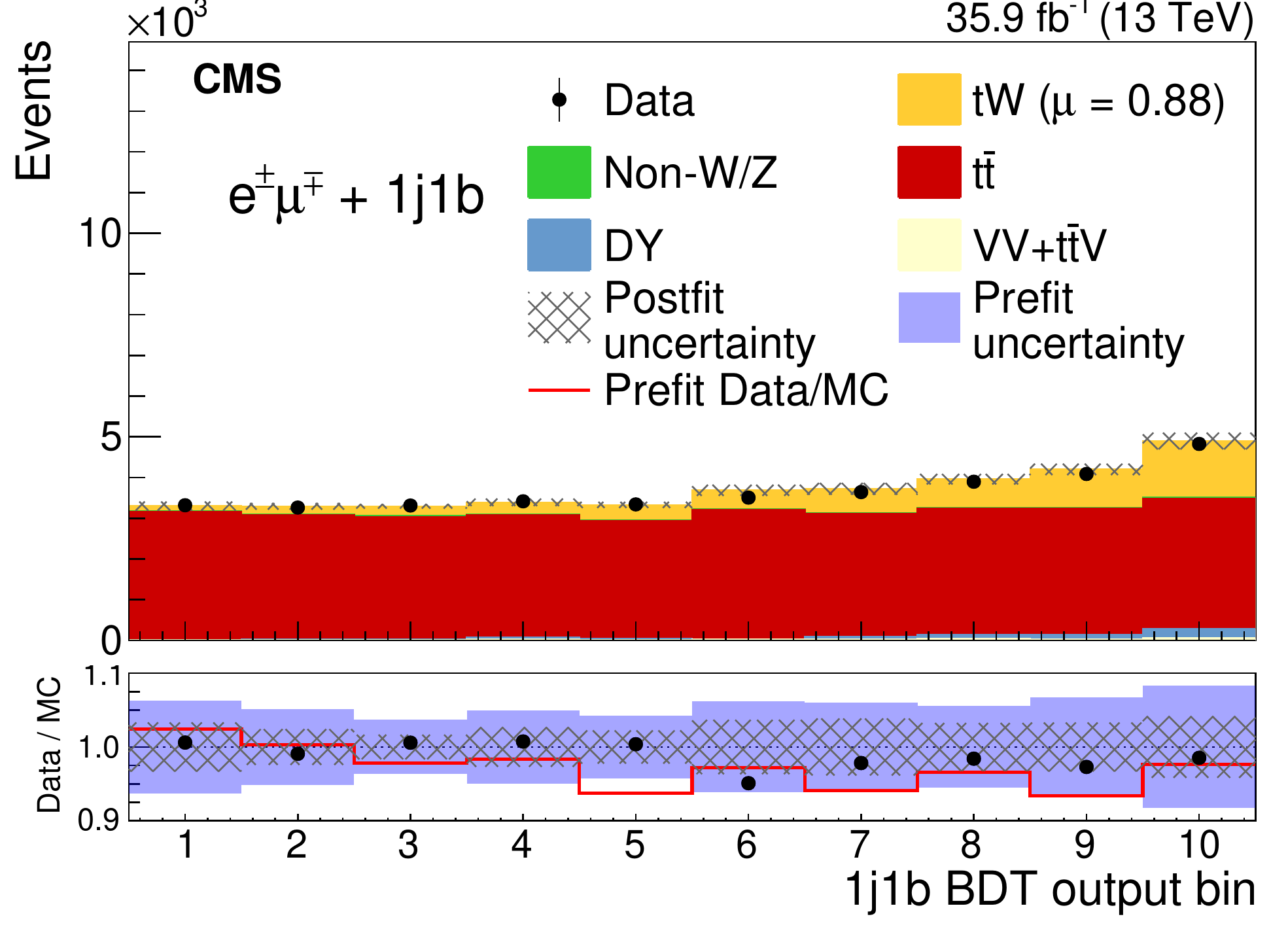}
\caption{}
\label{Fig-Wt-a}
\end{centering}
\end{subfigure} 
\begin{subfigure}{0.47\textwidth}
\begin{centering}
\includegraphics[scale=0.10]{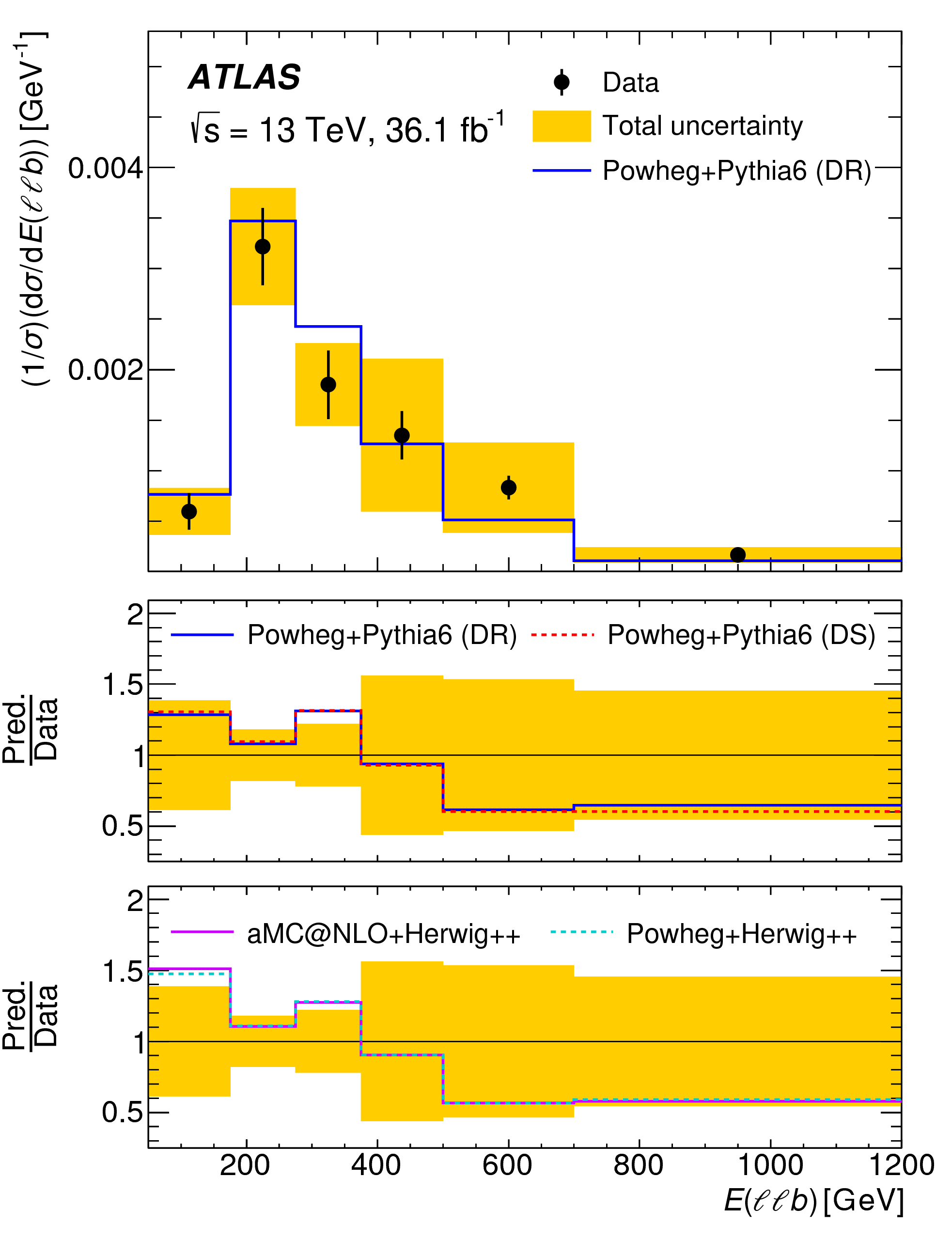}
\caption{}
\label{Fig-Wt-b}
\end{centering}
\end{subfigure} 
\caption{Representative figures from two recent measurements of associated $t\Wboson$ production at $\rts = 13$ \TeV. Shown are (a) the BDT output from the most sensitive channel in the CMS inclusive measurement which required exactly one reconstructed jet, which was $b$-tagged ($1j1b$)~\cite{CMS_tW}, and (b) the normalized differential measurement from the ATLAS result as a function of the $E_{\ell\ell b}$ variable described in the text~\cite{ATLAS_tW}.  The predicted signal yields for the CMS figure ($t\Wboson$) are scaled by a factor $\mu=0.88$ based on the results of a fit.  The various background processes are described in the reference.  The lower portions of the figures show the ratios of the measured data to the predictions from simulation, though in the case of (b) the predictions from simulation (pred.) are rather shown in the numerator.  The upper ratio plot for the ATLAS figure compares two separate simulated datasets which treat the interference between $\ttbar$ and $t\Wboson$ differently.  The lowermost ratio plot compares the measured data with separate predictions based on two generator models.  The shaded bands in all cases correspond to the total uncertainties for the given bin.}
\label{Fig-Wt}
\end{figure}

\subsection{Single top $t$-channel production}

The dominant leading-order production mode for single top or antitop quarks at the LHC is via a $t$-channel diagram.  A signature of $t$-channel production is the presence of an additional so-called spectator quark which is often produced in the more forward regions of the detector; the requirement of a reconstructed forward jet in such analyses is therefore helpful in identifying candidate signal events.  An inclusive cross-section measurement of this process was performed by CMS based on 35.9 \ifb \ of $\rts=13 \ \TeV$ Run 2 data \cite{CMS_t_chan}.  The production cross section was measured separately for top and antitop quarks ($\sigma_t$ and $\sigma_{\bar{t}}$) as well as their ratio ($\sigma_t/\sigma_{\bar{t}}$) where the two differ due to the partonic structure of the protons involved in the interactions, as governed by the parton distribution functions (PDFs).  The analysis targeted (anti)top quarks which decayed via $t\to\Wboson b\to\ell\nu_{\ell}b$, and where the event is attributed to a top or antitop quark according to the charge of the lepton. The fit used to extract the measurements of the respective cross sections was performed on a total of 12 BDT distributions, split according to the reconstructed lepton charge, the lepton flavour (either a reconstructed electron or muon) and different jet or $b$-jet multiplicities.  Such an approach allows for signal regions to be exploited with differing background compositions; in the case of the lepton charge for example certain processes ($t$-channel, $\Wboson+$jets) have charge-dependent yields, whereas others ($\ttbar$, $t\Wboson$, or multijet final states from various QCD processes) are largely charge-independent.  This feature can be observed in Figure~\ref{Fig-CMS-single-top-incl-a} which shows the yields for stacked SM predictions compared with those from the measured data in events where the reconstructed lepton was identified as a muon (the $\mu+$ jets channel).  The BDT from the subset of events corresponding to the $\mu^+$ selection with exactly two reconstructed jets, of which one is $b$-tagged, is shown in Figure~\ref{Fig-CMS-single-top-incl-b}.  The measured values of $\sigma_{t} = 126 \ \pm \ 1 \ {\rm (stat.)} \ \pm \ 22 \ {\rm (syst.)}$ pb, $\sigma_{\bar{t}} = 82 \ \pm \ 1 \ {\rm (stat.)} \ \pm \ 14 \ {\rm (syst.)}$ pb, and $\sigma_{t}/\sigma_{\bar{t}} = 1.66 \ \pm \ 0.02 \ {\rm (stat.)} \ \pm \ 0.05 \ {\rm (syst.)}$ are in agreement with a SM prediction calculated at NLO accuracy in QCD using \Hathor~\cite{Hathor1,Hathor2}.  The substantial reduction of the overall systematic uncertainty in the measurement of $\sigma_{t}/\sigma_{\bar{t}}$, which results from a partial cancellation of systematic uncertainties that affect both the numerator and denominator, highlights the advantage of taking the ratio.


A recent differential cross-section measurement of $t$-channel production was also performed by CMS with the same 35.9 \ifb \ of $\rts=13 \ \TeV$ data based on a similar set of selection criteria~\cite{CMS_t_chan_diff}.   The measured cross sections were presented as a function of a number of kinematic variables.  One example is the variable $\cos\theta^{\star}_{\rm pol.} = \frac{\vec{p}_{q'}^{(\rm top)}\cdot \vec{p}_{\ell}^{(\rm top)}}{|\vec{p}_{q'}^{(\rm top)}|\cdot |\vec{p}_{\ell}^{(\rm top)}|}$ -- the scalar product of the lepton and the spectator quark momenta, divided by the product of their magnitudes.  The superscripts in the expression denote that the momenta are evaluated in the (anti)top quark rest frame.  The $\cos\theta^{\star}_{\rm pol.}$ variable was chosen since it is sensitive to the top quark spin asymmetry $A_{\ell}$~\cite{CMS_t_chan_diff_asym}, and an extraction of that parameter from a fit to the measured data for the combined $e+\mu$ channels\footnote{The $e$ ($\mu$) channel corresponds to cases where the flavour of the reconstructed lepton ($\ell$) attributed to the (anti)top quark decay, $t\to b\Wboson \to b\ell\nu_{\ell}$, is an electron (muon).} yields a value of $A_{e+\mu} = 0.439 \ \pm \ 0.062$ in agreement with a SM prediction of 0.436 based on \PowhegBox \ at NLO.  The total uncertainty on the measured value quoted in the publication incorporates both statistical and systematic error sources, and is dominated by theoretical uncertainties.  

\begin{figure}[!htbp]
\begin{subfigure}{0.46\textwidth}
\begin{centering}
\includegraphics[scale=0.43]{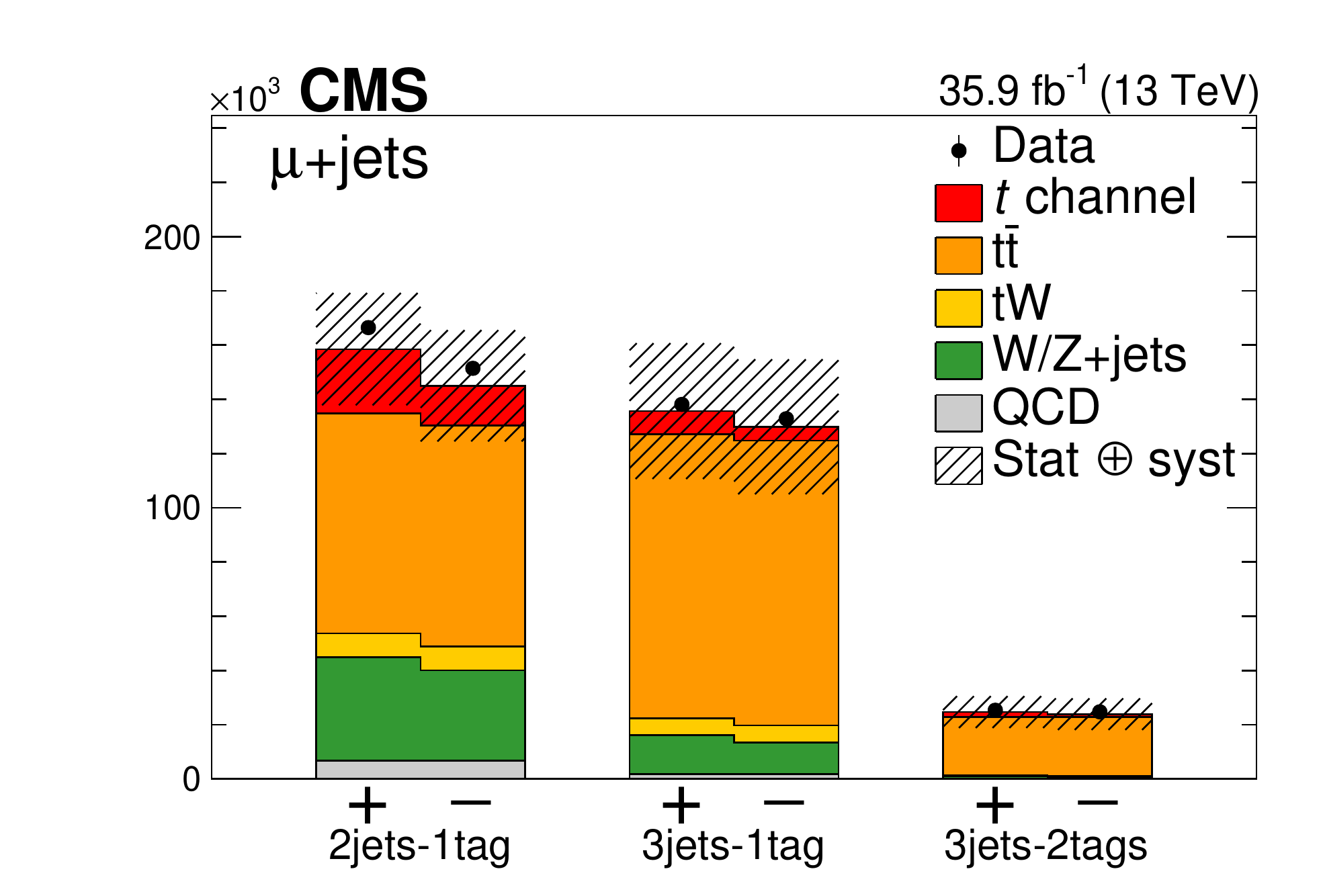}
\caption{}
\label{Fig-CMS-single-top-incl-a}
\end{centering}
\end{subfigure} 
\hspace{0.5cm}
\begin{subfigure}{0.46\textwidth}
\begin{centering}
\includegraphics[scale=0.28]{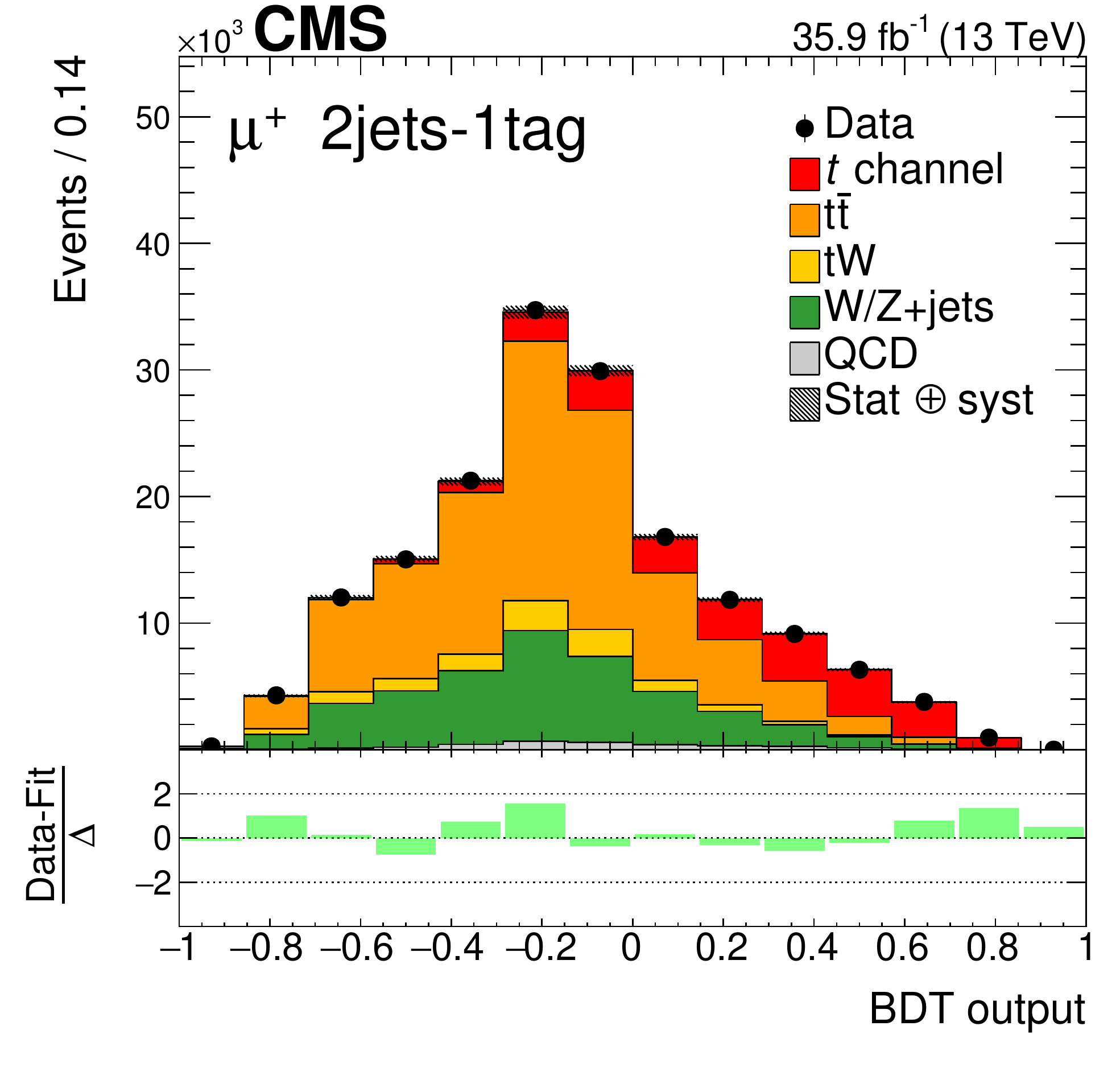}
\caption{}
\label{Fig-CMS-single-top-incl-b}
\end{centering}
\end{subfigure} 
\caption{Two figures from the recent CMS measurement of the inclusive $t$-channel single-top production cross section~\cite{CMS_t_chan}.  The selected figures correspond only to the $\mu$+jets channel and show (a) the decomposition of the event yields from data and prediction separated by lepton charge and jet multiplicity, and (b) the BDT output for the selection requiring one isolated muon and two hadronic jets, one of which is required to be $b$-tagged.  The lower portion of (b) shows bin-by-bin values of a pull quantity defined to be the difference between the measured yields in data and those from the prediction, divided by the uncertainty ($\Delta$), as described in the reference.}
\label{Fig-CMS-single-top-incl}
\end{figure}

\begin{figure}[!htbp]
\begin{subfigure}{0.47\textwidth}
\begin{centering}
\includegraphics[scale=0.63]{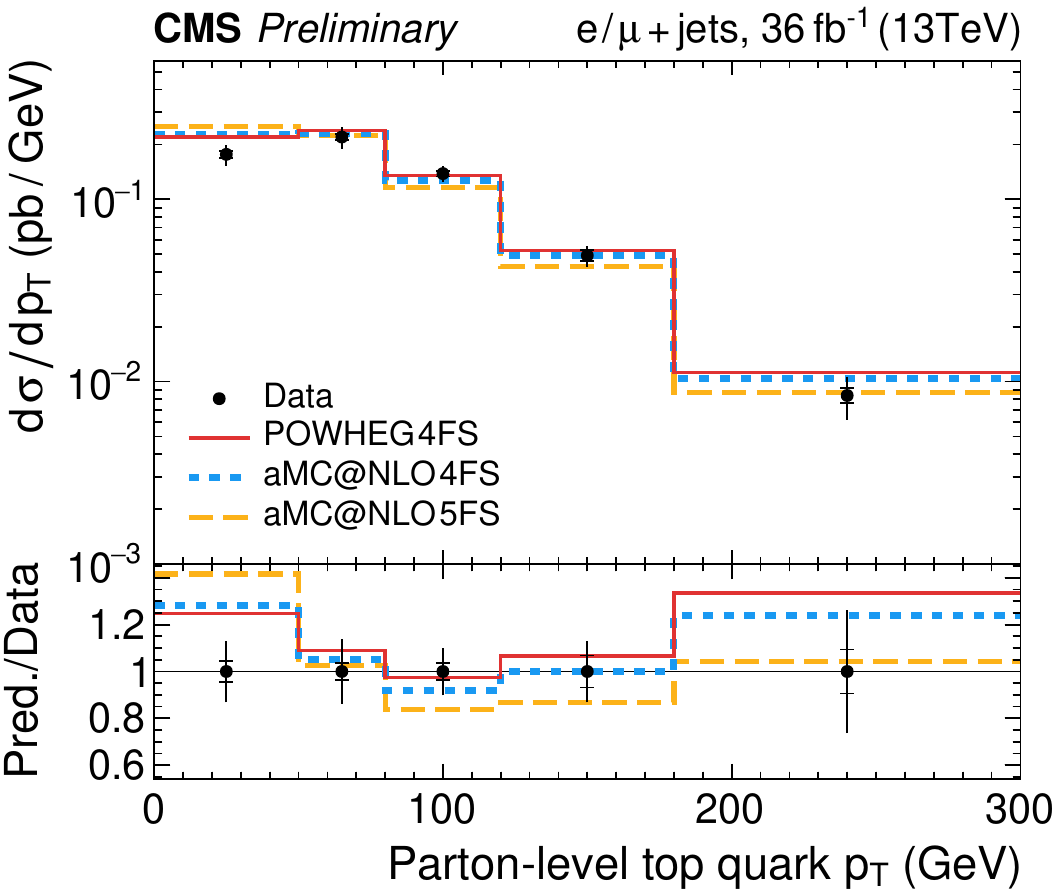}
\caption{}
\label{Fig-CMS-single-top-diff-a}
\end{centering}
\end{subfigure} 
\hspace{0.5cm}
\begin{subfigure}{0.47\textwidth}
\begin{centering}
\includegraphics[scale=0.63]{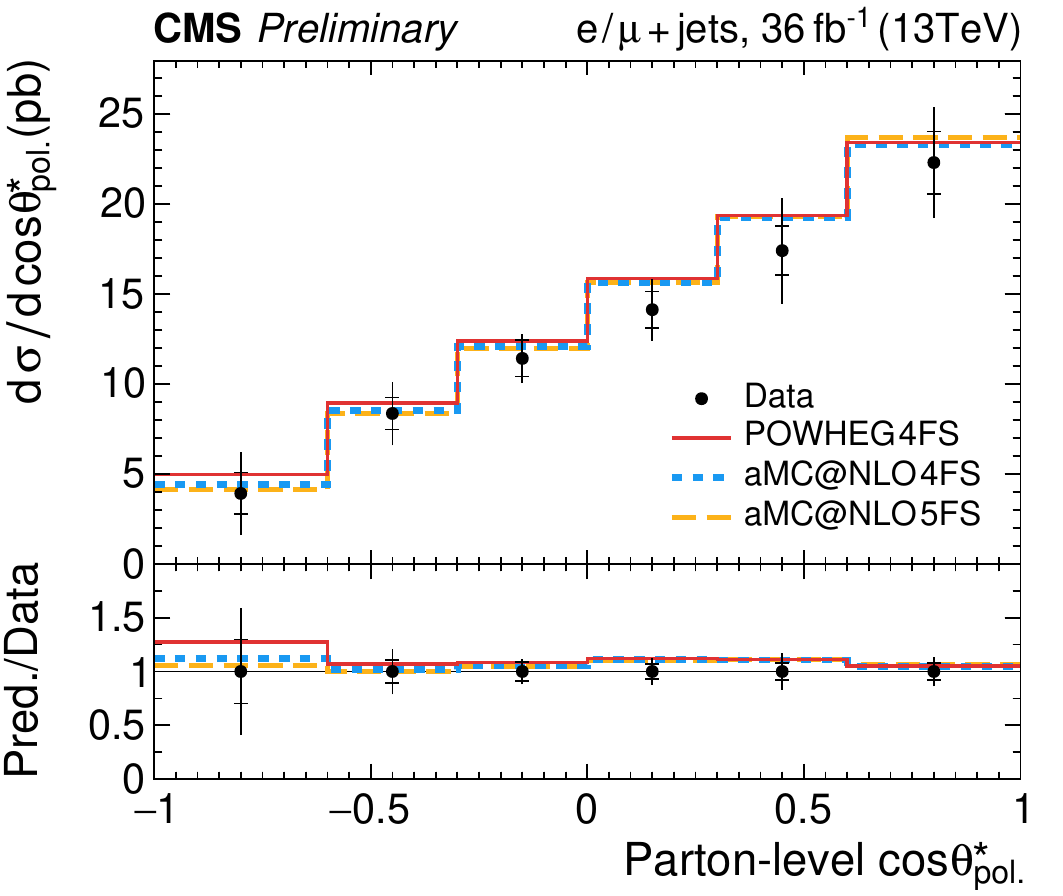}
\caption{}
\label{Fig-CMS-single-top-diff-b}
\end{centering}
\end{subfigure} 
\caption{Two differential cross-sections results from the CMS measurement of $t$-channel single top quark production~\cite{CMS_t_chan_diff}.  Shown are the absolute differential cross section measurements as a function of (a) the top quark \pt, and (b) the $\cos\theta^{\star}_{\rm pol.}$ variable described in the text.  Ratios of the values from various predictions to those from measured data are shown in the lower portions.}
\label{Fig-CMS-single-top-diff}
\end{figure}

Two representative absolute differential measurements are shown in Figures~\ref{Fig-CMS-single-top-diff-a} and \ref{Fig-CMS-single-top-diff-b} of the top quark transverse momentum and of the $\cos\theta^{\star}_{\rm pol.}$ variable, respectively.  In both cases the results correspond to parton-level quantities following the unfolding procedure described in the text~\cite{CMS_t_chan_diff}.  Based on the full set of differential results, the two predictions based on a four-flavour scheme (4FS, solid red and dashed blue lines) were found to agree better overall with the measured data than those based on a five-flavour scheme (5FS, dashed yellow lines), though as pointed out in the reference even for the 4FS a slight deviation can be seen for low top quark \pT.

A novel approach in this analysis was to measure the ratio of the production rates of top quarks to that of either top or antitop quarks $\frac{\sigma_{t}}{\sigma_{t}+\sigma_{\bar{t}}}$ differentially as a function of several variables, thereby providing a direct probe of the proton PDFs.  No tension was observed between the measured data and the predictions based on any of the considered PDF sets (NNPDF 3.0~\cite{CMS_t_chan_diff_PDF1}, MMHT 14~\cite{CMS_t_chan_diff_PDF2}, CT~10~\cite{CMS_t_chan_diff_PDF3}).



\subsection{Run 1 ATLAS + CMS combination of single top quark measurements}

A recent set of combinations of $\rts=7$ and $8 \ \TeV$ Run 1 measurements of single top quark production cross-sections were performed in order to enhance the precision with respect to the individual ATLAS and CMS measurements \cite{CMS_ATLAS_single_top}.  The combinations were performed separately for each centre-of-mass energy and production mode with the exception of the $s$-channel at $\rts=7 \ \TeV$ -- a total of five measurements per experiment in all.  A careful and detailed assessment of the individual sources of both the uncertainties and their correlations between the two experiments was crucial in order to be able to perform a meaningful combination.  The individual cross-section measurements were also used to probe the $V_{tb}$ vertex directly, by exploiting the fact that their ratio to the corresponding theoretical predictions (for which it is assumed that $V_{tb}=1$) is proportional to $|f_{\rm LV}V_{tb}|^2$, where $f_{LV}$ is a left-handed form factor which for the SM is assumed to be real and to have the value $f_{\rm LV}=1$~\cite{CMS_ATLAS_single_top_theory}.  Since this direct relationship is common to all measurements, the $|f_{\rm LV}V_{tb}|$ values were able to be combined, both individually for each centre-of-mass value and production mode, but also based on all measurements\footnote{In the final combination the CMS $s$-channel measurement at $\rts=8 \ \TeV$ was excluded for the reasons detailed in the publication.}.  The final relative precision of $\frac{\delta|f_{\rm LV}V_{tb}|}{|f_{\rm LV}V_{tb}|} \approx 3.7\%$, dominated by the most sensitive $t$-channel measurements, is the most precise direct measurement for the given set of assumptions\footnote{It is assumed that $|V_{td}|,|V_{ts}| \ll |V_{tb}|$ and that the $Wtb$ vertex features a left-handed coupling as in the SM.  No assumptions are made on the number of quark generations nor on CKM unitarity.}.  Figures~\ref{Fig-LHC-single-top-a} and \ref{Fig-LHC-single-top-b} show the summary of the individual cross-section measurements and their combinations, and the per-channel and overall combination results for $|f_{\rm LV}V_{tb}|$, respectively.

\begin{figure}[!htbp]
\begin{subfigure}{0.47\textwidth}
\begin{centering}
\includegraphics[scale=0.375]{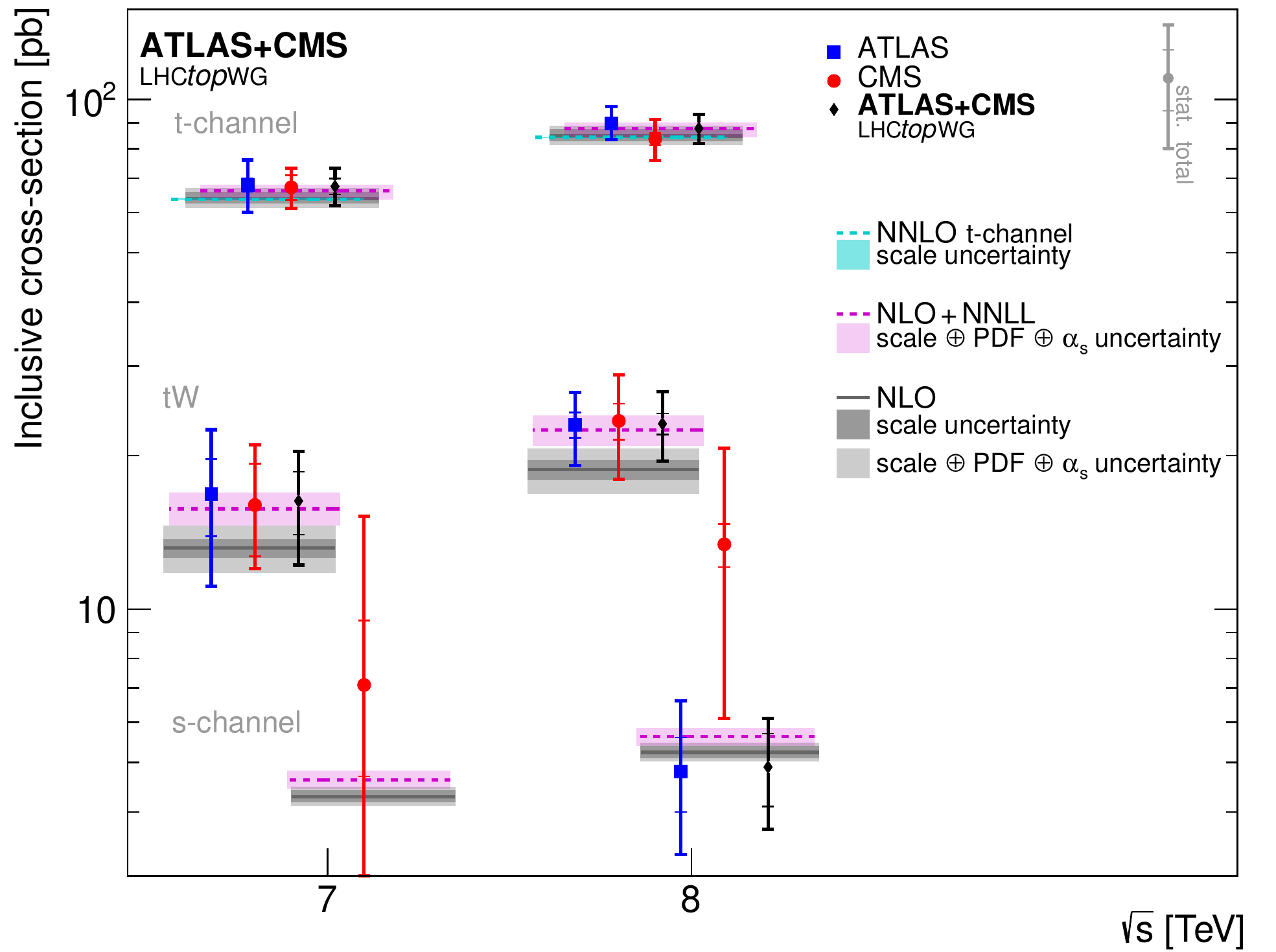}
\caption{}
\label{Fig-LHC-single-top-a}
\end{centering}
\end{subfigure} 
\hspace{0.5cm}
\begin{subfigure}{0.47\textwidth}
\begin{centering}
\includegraphics[scale=0.31]{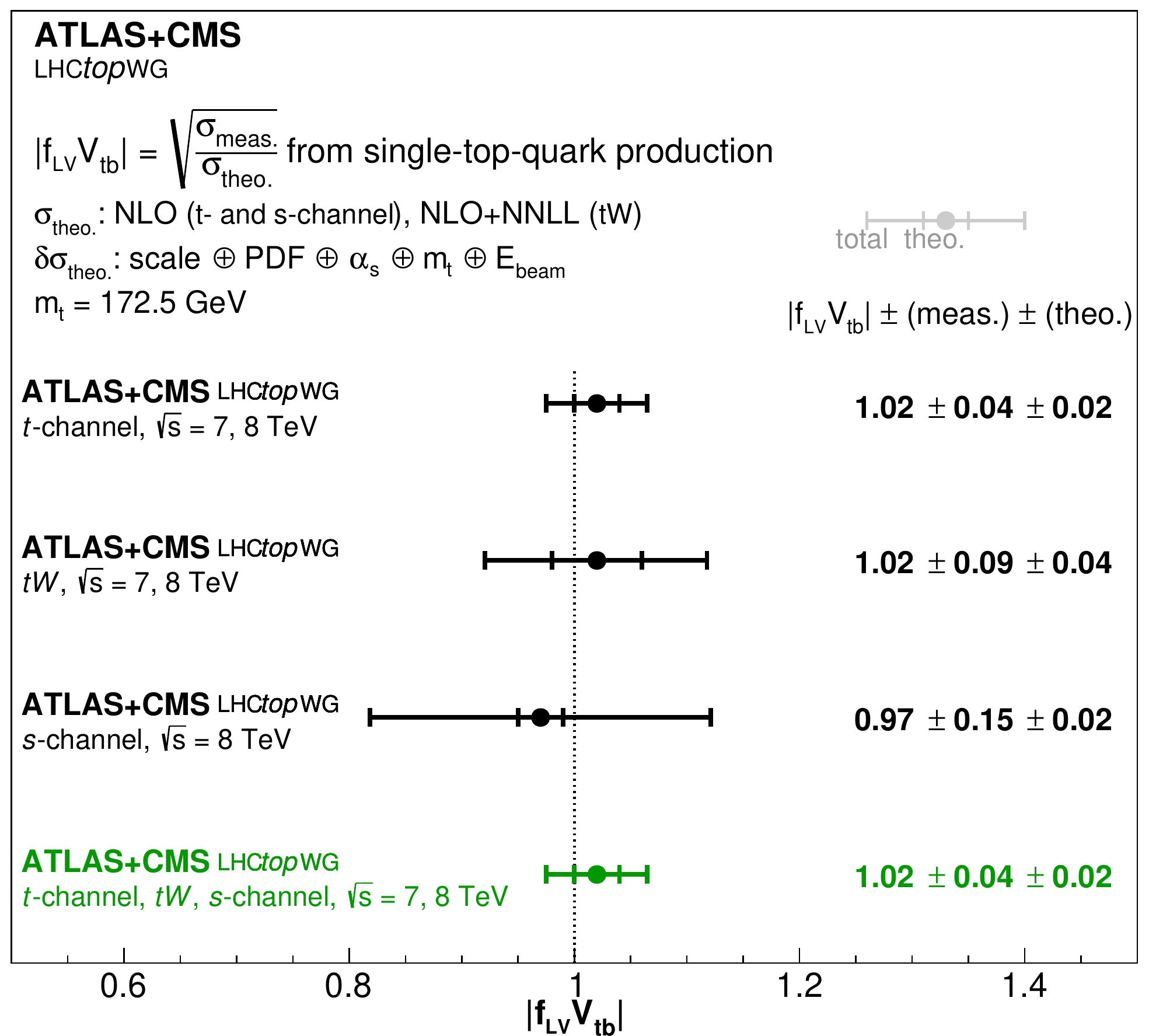}
\caption{}
\label{Fig-LHC-single-top-b}
\end{centering}
\end{subfigure} 
\caption{Summary plots from the recent combination of $\rts = $7 and 8 \TeV \ results showing (a) individual per-channel single top quark production measurements from ATLAS and CMS, and (b) the full combination of $|f_{\rm LV}V_{tb}|$ measurements~\cite{CMS_ATLAS_single_top}.}
\label{Fig-LHC-single-top}
\end{figure}


\section{Additional top quark measurements sensitive to modelling and non-SM effects}

\subsection{$\ttbar$ + $t\Wboson b$ interference}

A new ATLAS result compares different ways to treat the interference between singly and doubly resonant $\Wplus b\Wminus\bar{b}$ production based on 36.1 \ifb \ of $\rts=13 \ \TeV$ data~\cite{ATLAS_tt_Wt_QI}.  An event selection was applied to target either $\ttbar$ or $t\Wboson$ production with opposite-sign dileptonic final states ($ee,e\mu,\mu\mu$).  The analysis makes use of the quantity



$$m_{b\ell}^{\rm minimax} \equiv \min \{\max \{m_{b_{1}\ell_{1}},m_{b_{2}\ell_{2}}\}, \max \{m_{b_{1}\ell_{2}},m_{b_{2}\ell_{1}}\}\}, $$

\noindent which was constructed in order to heighten the differences between the DR and DS predictions described for the earlier ATLAS $t\Wboson$ analysis.  The subscripts in the invariant $b\ell$ mass terms in the quantity denote the choice of $\ell$-$b$ association based on the combinatorics, since one does not know {\it a priori} which reconstructed lepton to associate with which reconstructed $b$-jet.  Figure~\ref{Fig-ATLAS-QI-a} shows the distribution of the $m_{b\ell}^{\rm minimax}$ variable in measured data compared with the stacked SM prediction, where the greatest sensitivity to the interference effects is exhibited in the uppermost bins.   

\begin{figure}[!htbp]
\begin{subfigure}{0.47\textwidth}
\begin{centering}
\includegraphics[scale=0.37]{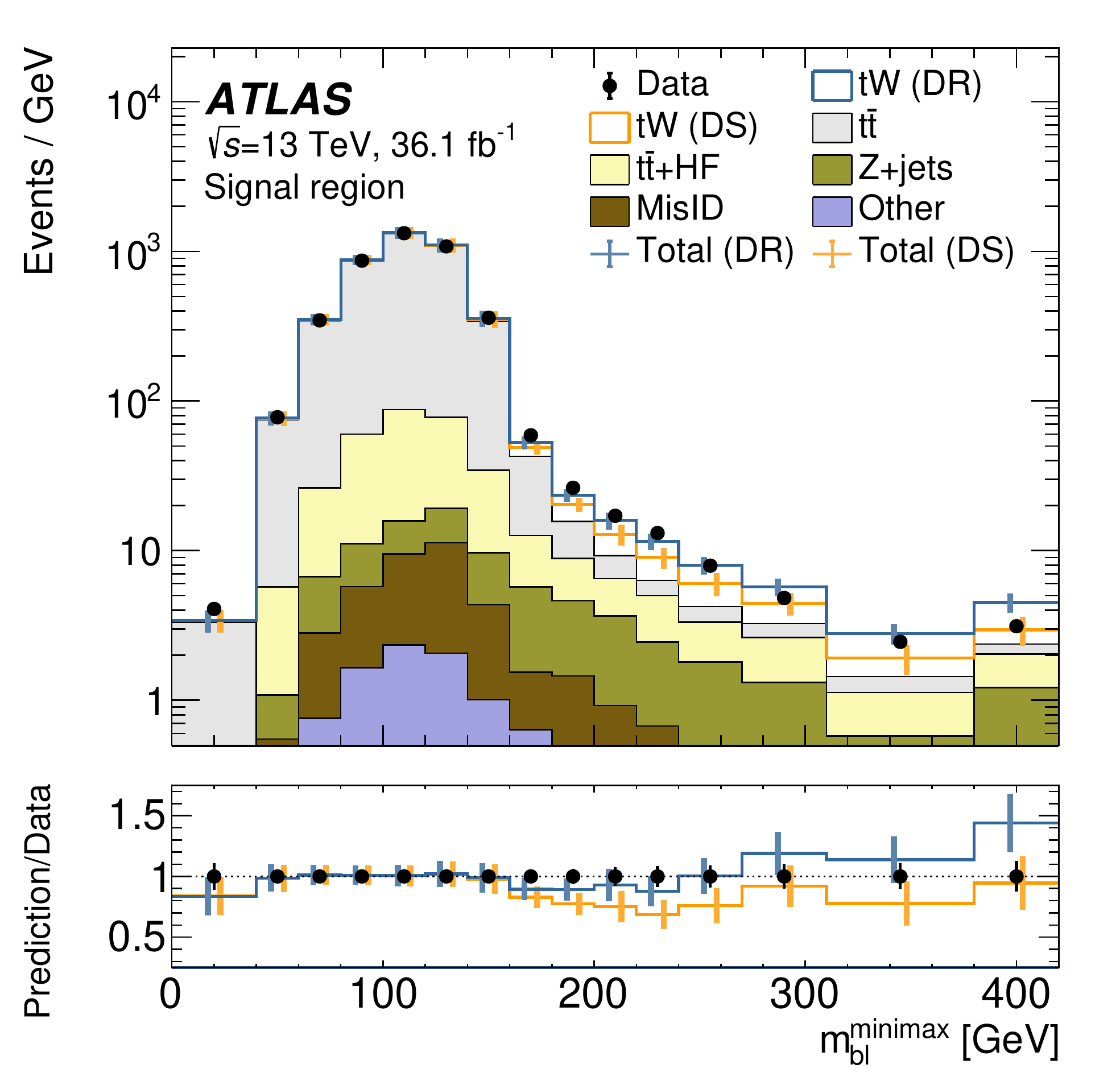}
\caption{}
\label{Fig-ATLAS-QI-a}
\end{centering}
\end{subfigure} 
\hspace{0.5cm}
\begin{subfigure}{0.47\textwidth}
\begin{centering}
\includegraphics[scale=0.35]{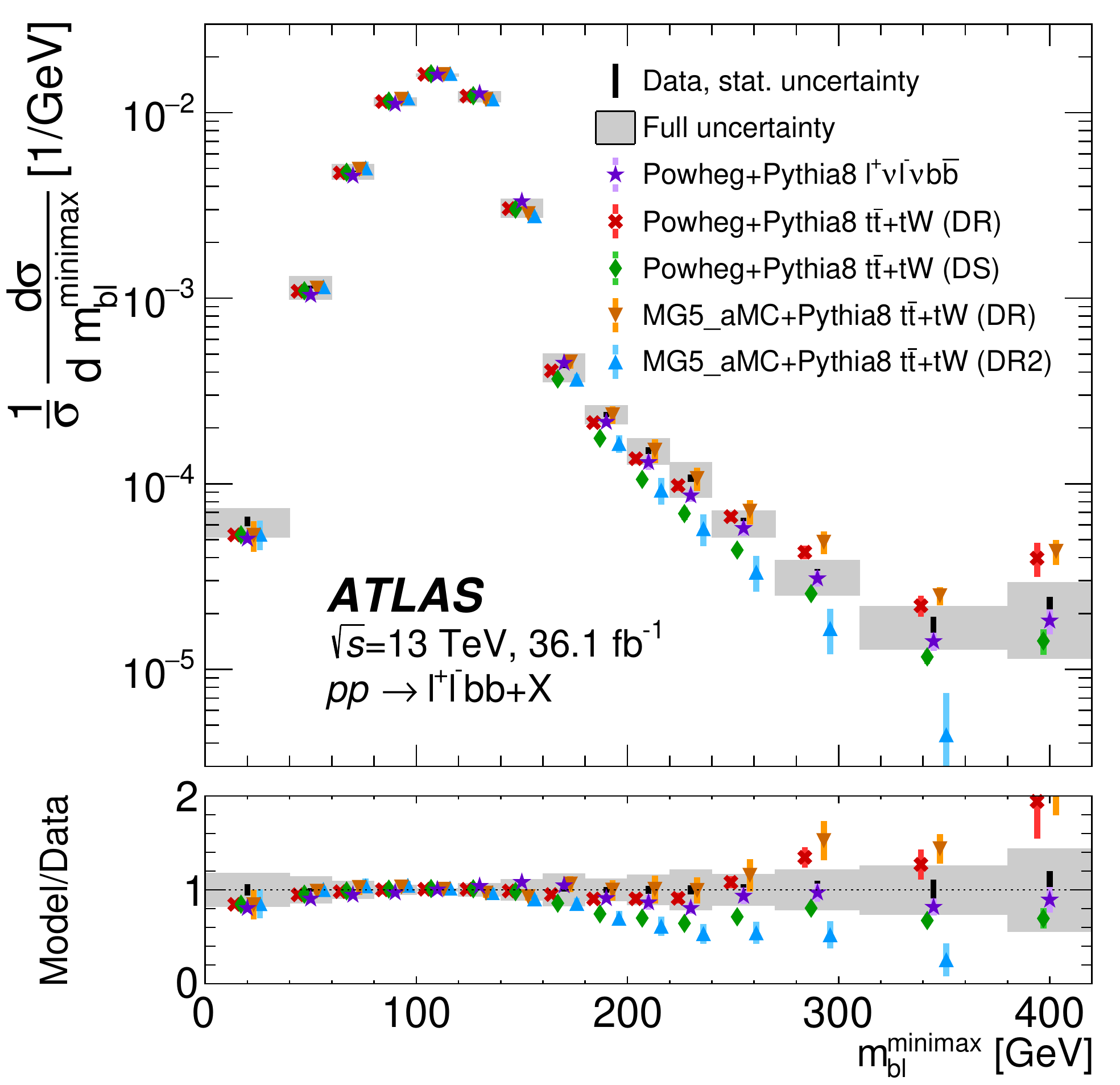}
\caption{}
\label{Fig-ATLAS-QI-b}
\end{centering}
\end{subfigure} 
\caption{Figures from the $\rts=13 \ \TeV$ ATLAS measurement to investigate the treatment of interference between $\Wboson t$ and $\ttbar$ processes~\cite{ATLAS_tt_Wt_QI}.  Shown are (a) distributions comparing the measured data to stacked SM predictions for a variable sensitive to the interference effects in the tails, and (b) the normalized cross-section measurement as a function of the same variable.  The values shown in the lower portion of (a) correspond to the sum of the predicted SM yields where the $\Wboson t$ contribution was based either on the DR (blue) or DS (orange) prediction.  In (b) the values correspond to the ratio of the given signal $\Wboson t$ prediction (shown for the various coloured markers) relative to that from the measured data.}
\label{Fig-ATLAS-tt-Wt-QI}
\end{figure}

The associated normalized differential cross-section measurement is shown in Figure~\ref{Fig-ATLAS-QI-b}, where one can compare the measured results from data with the various predictions. The DR and DS predictions were seen to differ from each other significantly but each to be within $2\sigma$ of the measured data.  In contrast, good agreement was observed between the data and the full NLO \PowhegBoxRes \ prediction~\cite{ATLAS_tt_Wt_QI_prediction}, which properly incorporates the effects due to the interference.  The result highlights the benefits to be gained from a proper treatment of interference; employing such simulations in future results, where available, will prove useful in greatly reducing the associated systematic uncertainties.


\subsection{Search for non-SM physics in $\ttbar$ and $t\Wboson$ events}

A recent CMS result targeted a selection of data events dominated by $\ttbar$ and $t\Wboson$ processes with opposite-sign dileptonic final states in order to search for evidence of new physics \cite{CMS_tt_dilep_new_phys}.  The approach is based on the fact that higher-order, non-SM couplings of the top quark to gluons or to heavy electroweak bosons ($\Wboson,\Zboson$) are expected to modify both the production rates and kinematic distributions for $\ttbar$ and $t\Wboson$, and crucially by amounts which differ for the two processes.  This sensitivity was exploited by constructing and training dedicated neural networks to improve the ability to discriminate between signal and background.  The designation of which process(es) correspond to the signal or background varies depending on the context, but typically the SM $\ttbar$ and $t\Wboson$ processes are treated collectively as background in order to heighten the sensitivity to non-SM effects\footnote{In the case of searching for flavour-changing neutral-current (FCNC) effects, the $t\Wboson$ process is rather interpreted as the signal and the $\ttbar$ as the background.}.  The results were interpreted in the context of SM EFT.  

\begin{figure}[!htbp]
\begin{subfigure}{0.47\textwidth}
\begin{centering}
\includegraphics[scale=0.35]{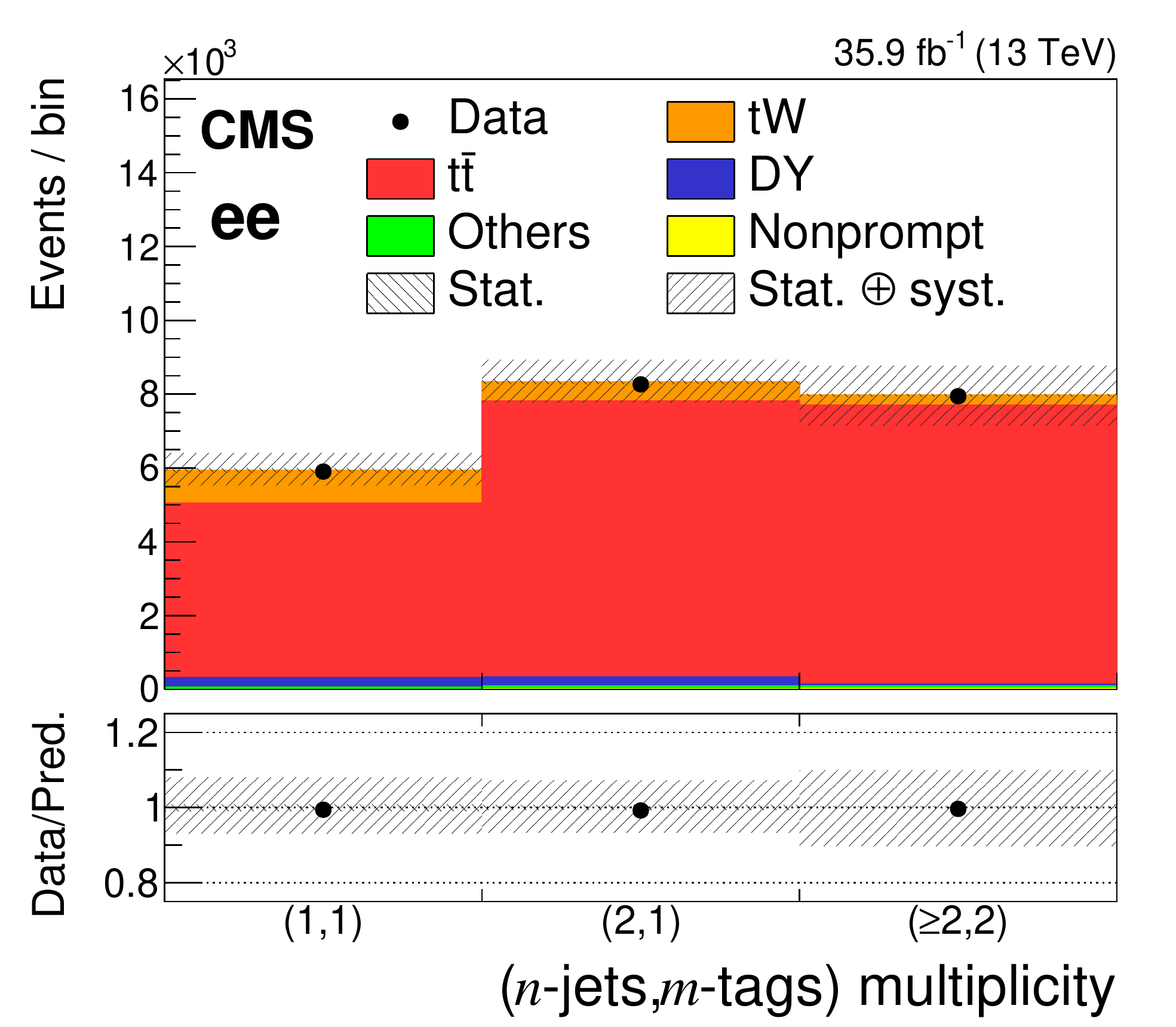}
\caption{}
\label{Fig-CMS-tt-tW-EFT-a}
\end{centering}
\end{subfigure} 
\hspace{0.5cm}
\begin{subfigure}{0.47\textwidth}
\begin{centering}
\includegraphics[scale=0.35]{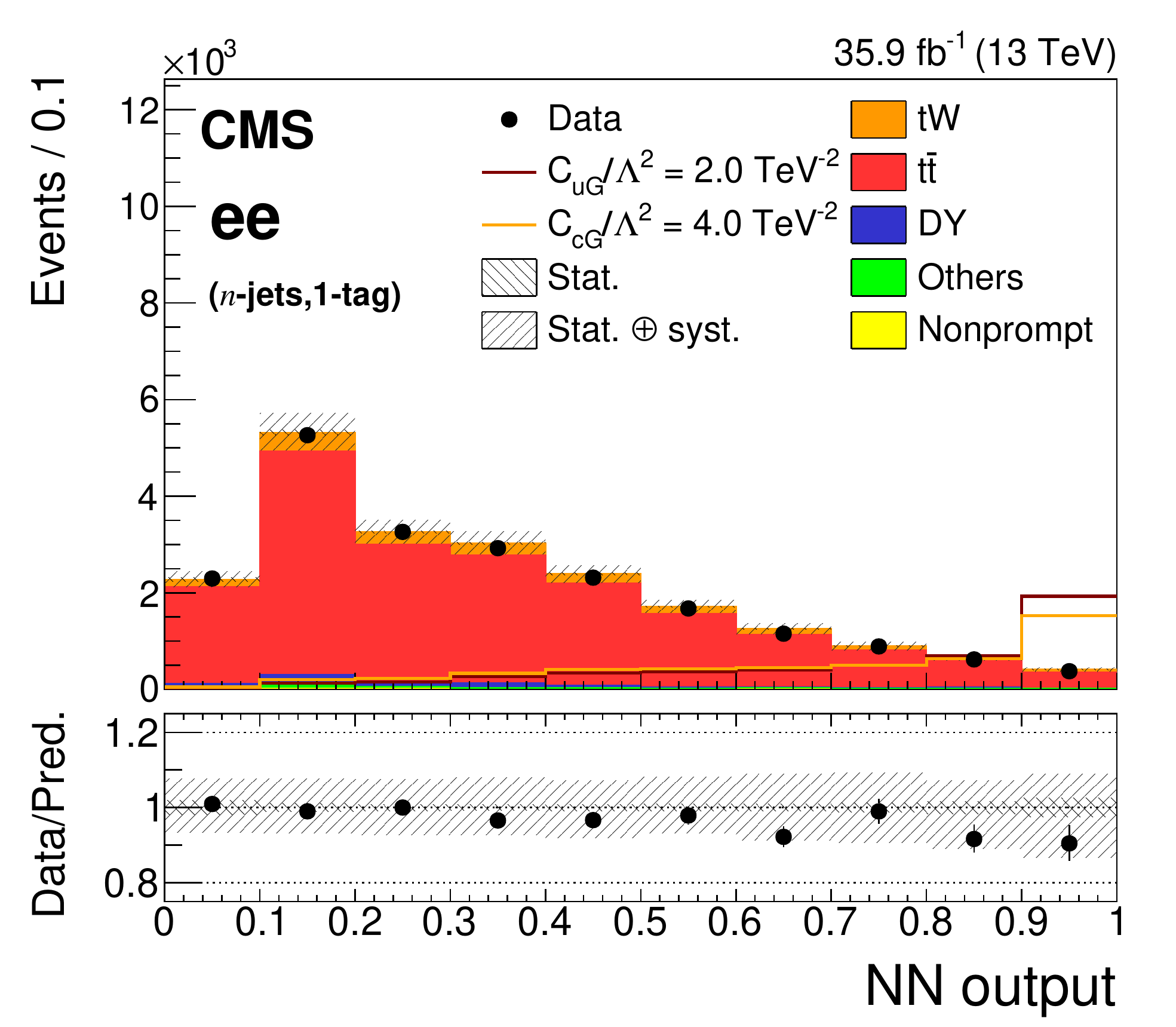}
\caption{}
\label{Fig-CMS-tt-tW-EFT-b}
\end{centering}
\end{subfigure} 
\caption{(a) Comparison of event yields from data and SM predictions for three separate jet and $b$-jet multiplicity bins, and (b) the NN output in the 1-$b$-tag category for the $ee$ channel from the CMS $\rts=13 \ \TeV$ search for new physics in dileptonic  $\ttbar$ and $t\Wboson$ final states~\cite{CMS_tt_dilep_new_phys}.  The various SM processes shown in the stacked distribution, including those from non-prompt leptons and Drell-Yan (DY) production, are described in the reference.  The lower portions of the figures show bin-by-bin the ratio of the measured yields in data to the sums from the various SM predictions.}
\label{Fig-CMS-tt-tW-EFT}
\end{figure}

The distributions in Figure~\ref{Fig-CMS-tt-tW-EFT-a}, separated into different jet and $b$-jet multiplicity bins, give a sense of the relative yields from the two processes specifically for the selection featuring two reconstructed electrons ($ee$ channel).  The inclusive $n$-jet distribution of the NN output for the selection requiring a single $b$-jet (1-tag), also in the $ee$ channel\footnote{The two lepton flavours ($ee$ in this case) refer to those previously described for the $t\Wboson$ cross-section measurements.}, is shown in Figure~\ref{Fig-CMS-tt-tW-EFT-b}.  Here one can note the contrast between the shapes from the combined SM background, which in this case includes both $\ttbar$ and $t\Wboson$, and the lines for two representative values of different non-SM couplings relative to an energy scale ($\Lambda$) squared.  Binned likelihood fits were performed for the various operators by scanning, one coefficient at a time, through non-zero values of the given effective coupling.  No significant deviations from the SM were observed, but the measurement allowed for a stringent set of new limits to be placed on the Wilson coefficients from non-SM dimension-six operators.

\subsection{Associated $t+\Zboson$ production}

Searches for the associated production of a single top or antitop quark and a $\Zboson$ boson have been performed both by CMS and ATLAS \cite{CMS_tZq,ATLAS_tZq}, including the recent observation at $\rts=13 \ \TeV$ by CMS \cite{CMS_tZq_obs}.  Such searches would generally fit into the category of analyses presented in these proceedings, but were attributed to another presentation focusing on rare SM processes featuring top quarks.

\section{Summary}

A number of recent ATLAS and CMS measurements featuring top quark signatures have been presented in these proceedings.    Such measurements serve as rigorous tests of SM predictions related to the production and decay modes of the most massive SM particle.  The large proton-proton collision datasets collected during Run 2 at the LHC for both experiments have allowed not only for precise inclusive cross-section measurements, but also for differential measurements for a number of interesting processes such as $t$-channel single top, and associated $\Wboson t$, $\ttbar\Zboson$ or $\ttbar\gamma$ production.  A thorough combination of Run 1 ATLAS and CMS single top quark production cross-section measurements was performed, yielding improved precision with respect to individual measurements.  Measurements were also performed to search for possible signs of new physics in $\ttbar$ and $\Wboson t$ production, and to compare various ways to model the interference between singly and doubly resonant $\Wboson b\Wboson b$ production.  No significant deviations from SM predictions have been observed, but new limits were able to be placed on a number of parameters for non-SM models.





\begin{thebibliography}{99}

\bibitem{ATLASExperiment}
ATLAS Collaboration, \emph{The ATLAS Experiment at the CERN Large Hadron Collider}, \\ JINST {\bf 3} (2008) S08003 [\href{http://inspirehep.net/record/796888}{\tt inspirehep.net/record/796888}].

\bibitem{CMSExperiment}
CMS Collaboration, \emph{The CMS Experiment at the CERN LHC}, JINST {\bf 3} (2008) S08004 [\href{http://inspirehep.net/record/796887}{\tt inspirehep.net/record/796887}].

\bibitem{ATLAS_tty}
ATLAS Collaboration, \emph{Measurement of the inclusive and differential fiducial cross-section of $t\bar{t}\gamma$ in leptonic final states at $\sqrt{s}$ = 13 \TeV \ in ATLAS}, Eur. Phys. J. C {\bf 79} (2019) 382 [\href{https://arxiv.org/abs/1812.01697}{\tt arXiv:1605.04436v2}].

\bibitem{ATLAS_tty_NLO_predictions}
K. Melnikov, A. Scharf and M. Schulze, \emph{QCD corrections to top quark pair production in association with a photon at hadron colliders}, Phys. Rev. D {\bf 83} (2011) 074013, [\href{https://arxiv.org/abs/1102.1967}{\tt arXiv:1102.1967}].

\bibitem{ATLAS_aMCatNlo}
J. Alwall {\it et al.}, \emph{The automated computation of tree-level and next-to-leading order differential cross sections, and their matching to parton shower simulations}, JHEP {\bf 07} (2014) 079 [\href{https://arxiv.org/abs/1405.0301}{\tt arXiv:1405.0301}].

\bibitem{ATLAS_PowhegBox}
S. Frixione, P. Nason and C. Oleari, \emph{Matching NLO QCD computations with parton shower simulations: the POWHEG method}, JHEP {\bf 11} (2007) 070, [\href{https://arxiv.org/abs/0709.2092}{\tt arXiv:0709.2092}].

\bibitem{ATLAS_Pythia8}
T. Sj$\ddot{\textrm{o}}$strand {\it et al.}, \emph{An introduction to PYTHIA 8.2}, Comput. Phys. Commun. {\bf 191} (2015) 159, [\href{https://arxiv.org/abs/1410.3012}{\tt arXiv:1410.3012}].


\bibitem{CMS_ty}
CMS Collaboration, \emph{Evidence for the associated production of a single top quark and a photon in proton-proton collisions at $\sqrt{s}$ = 13 \TeV}, Phys. Rev. Lett. {\bf 121} (2018) 221802 [\href{https://arxiv.org/abs/1808.02913}{\tt arXiv:1808.02913v2}].






\bibitem{ATLAS_ttV}
ATLAS Collaboration, \emph{Measurement of the $t\bar{t}\Zboson$ and $t\bar{t}\Wboson$ cross sections in proton-proton collisions at $\sqrt{s}$ = 13 \TeV \ with the ATLAS detector}, Phys. Rev. D {\bf 99} (2019) 072009 [\href{https://arxiv.org/abs/1901.03584}{\tt arXiv:1901.03584v2}].

\bibitem{ATLAS_ttV_3p2}
ATLAS Collaboration, \emph{Measurement of the $\ttbar\Zboson$ and $\ttbar\Wboson$ production cross sections in multi lepton final states using 3.2 \ifb \ of pp collisions at $\sqrt{s}$ = 13 \TeV \ with the ATLAS detector}, Eur. Phys. J. C {\bf 77} (2017) 40 [\href{https://arxiv.org/abs/1609.01599}{\tt arXiv:1609.01599v2}].


\bibitem{ATLAS_ttV_theory2}
D. de Florian {\it et al.}, \emph{Handbook of LHC Higgs cross sections: 4. deciphering the nature of the Higgs sector}, CERN Report CERN-2017-002-M (2016) [\href{https://arxiv.org/abs/1610.07922}{\tt arXiv:1610.07922}].

\bibitem{CMS_ttZ_diff}
CMS Collaboration, \emph{Measurement of top quark pair production in association with a $\Zboson$ boson in proton-proton collisions at $\sqrt{s}$ = 13 \TeV}, (March 2019) [\href{http://cms-results.web.cern.ch/cms-results/public-results/preliminary-results/TOP-18-009/index.html}{\tt CMS-PAS-TOP-18-009 }].


\bibitem{CMS_ttZ_diff_EFT1}
C. Zhang and S. Willenbrock, \emph{Effective-field-theory approach to top-quark production and decay}, Phys. Rev. D {\bf 83} (2011) 034006 [\href{https://arxiv.org/abs/1008.3869}{\tt arXiv:1008.3869}].

\bibitem{CMS_ttZ_diff_EFT2}
G. Durieux, F. Maltoni, and C. Zhang, \emph{A global approach to top-quark flavor-changing interactions}, Phys. Rev. D {\bf 91} (2015) 074017 [\href{https://arxiv.org/abs/1412.7166}{\tt arXiv:1412.7166}].

\bibitem{CMS_ttZ_diff_EFT3}
B. Grzadkowski {\it et al.}, \emph{Dimension-Six Terms in the Standard Model Lagrangian}, JHEP {\bf 10} (2010) 085 [\href{https://arxiv.org/abs/1008.4884}{\tt arXiv:1008.4884}].

\bibitem{CMS_ttZ_diff_EFT4}
J. A. Aguilar Saavedra {\it et al.}, \emph{Effective-field-theory approach to top-quark production and decay}, LHC Top Working Group Note (2018) [\href{https://arxiv.org/abs/1802.07237}{\tt arXiv:1802.07237}].


\bibitem{CMS_ttV}
CMS Collaboration, \emph{Measurement of the cross section for top quark pair production in association with a $\Wboson$ of $\Zboson$ boson in proton-proton collisions at $\sqrt{s}$ = 13 \TeV}, JHEP {\bf 08} (2018) 011 [\href{https://arxiv.org/abs/1711.02547}{\tt arXiv:1711.02547v2}].

\bibitem{CMS_tW}
CMS Collaboration, \emph{Measurement of the production cross section for single top quarks in association with $\Wboson$ bosons in proton-proton collisions at $\sqrt{s}$ = 13 \TeV}, JHEP {\bf 10} (2018) 117 [\href{https://arxiv.org/abs/1805.07399}{\tt arXiv:1805.07399v2}].

\bibitem{CMS_tW_NNLO_prediction}
N. Kidonakis, \emph{Theoretical results for electroweak-boson and single-top production}, Proceedings from 23rd International Workshop on Deep-Inelastic Scattering and Related Subjects (DIS 2015), Dallas, Texas, USA, April 27 - May 01, 2015, Volume DIS2015, p. 170 [\href{https://arxiv.org/abs/1506.04072}{\tt arXiv:1506.04072}].

\bibitem{ATLAS_tW}
ATLAS Collaboration, \emph{Measurement of the differential cross-sections of a single top quark produced in association with a $\Wboson$ boson at $\sqrt{s}$ = 13 \TeV \ with ATLAS}, Eur. Phys. J. C {\bf 78} (2018) 186 [\href{https://arxiv.org/abs/1712.01602}{\tt arXiv:1712.01602v2}].

\bibitem{ATLAS_tW_DR_DS}
S. Frixione {\it et al.}, \emph{Single-top hadroproduction in association with a W boson}, JHEP {\bf 07} (2008) 029, [\href{https://arxiv.org/abs/0805.3067}{\tt arXiv:0805.3067}].


\bibitem{ATLAS_tW_Herwigpp}
G. Corcella {\it et al.}, \emph{HERWIG 6.5: an event generator for Hadron Emission Reactions With Interfering Gluons (including supersymmetric processes)}, JHEP {\bf 01} (2001) 010 [\href{https://arxiv.org/abs/hep-ph/0011363}{\tt arXiv:0011363}]

\bibitem{CMS_t_chan}
CMS Collaboration, \emph{Measurement of the single top quark and antiquark production cross sections in the t channel and their ratio in proton-proton collisions at $\sqrt{s}$ = 13 \TeV}, Submitted to Phys. Lett. B (December 2018) [\href{https://arxiv.org/abs/1812.10514}{\tt CMS-TOP-17-011}].

\bibitem{Hathor1}
M. Aliev {\it et al.}, \emph{HATHOR: HAdronic Top and Heavy quarks crOss section calculatoR}, Comput. Phys. Commun. {\bf 182} (2011) 1034 [\href{https://arxiv.org/abs/1007.1327}{\tt arXiv:1007.1327}].

\bibitem{Hathor2}
P. Kant {\it et al.}, \emph{HATHOR for single top-quark production: Updated predictions and uncertainty estimates for single top-quark production in hadronic collisions}, Comput. Phys. Commun. {\bf 191} (2015) 74 [\href{https://arxiv.org/abs/1406.4403}{\tt arXiv:1406.4403}].

\bibitem{CMS_t_chan_diff}
CMS Collaboration, \emph{Measurement of differential cross sections and charge ratios for t-channel single top quark production at 13 \TeV}, (March 2019) [\href{http://cms-results.web.cern.ch/cms-results/public-results/preliminary-results/TOP-17-023/index.html}{\tt CMS-TOP-17-023}].

\bibitem{CMS_t_chan_diff_asym}
J. A. Aguilar-Saavedra and J. Bernabeu, \emph{W polarisation beyond helicity fractions in top quark decays}, Nucl. Phys. B {\bf 840} (2010) 349 [\href{https://arxiv.org/abs/1005.5382}{\tt arXiv:1005.5382}].

\bibitem{CMS_t_chan_diff_PDF1}
NNPDF Collaboration, \emph{Parton distributions for the LHC Run II}, JHEP {\bf 04} (2015) 040 [\href{https://arxiv.org/abs/1410.8849}{\tt arXiv:1410.8849}].

\bibitem{CMS_t_chan_diff_PDF2}
L. A. Harland-Lang {\it et al.}, \emph{Parton distributions in the LHC era: MMHT 2014 PDFs}, Eur. Phys. J. C {\bf 75} (2015) 204 [\href{https://arxiv.org/abs/1412.3989}{\tt arXiv:1412.3989}].

\bibitem{CMS_t_chan_diff_PDF3}
H.-L. Lai {\it et al.}, \emph{New parton distributions for collider physics}, Phys. Rev. D {\bf 82} (2010) 074024 [\href{https://arxiv.org/abs/1007.2241}{\tt arXiv:1007.2241}].

\bibitem{CMS_ATLAS_single_top}
ATLAS \& CMS Collaborations, \emph{Combinations of single-top-quark production cross-section measurements and $|f_{\mathrm{LV}}V_{tb}|$ determinations at $\sqrt{s}$ = 7 and 8 \TeV \ with the ATLAS and CMS experiments}, JHEP {\bf 05} (2019) 088 [\href{https://arxiv.org/abs/1902.07158}{\tt arXiv:1902.07158 }].

\bibitem{CMS_ATLAS_single_top_theory}
J. A. Aguilar-Saavedra, \emph{A minimal set of top anomalous couplings}, Nucl. Phys. B {\bf 812} (2009) 181 [\href{https://arxiv.org/abs/0811.3842}{\tt arXiv:0811.3842}].

\bibitem{ATLAS_tt_Wt_QI}
ATLAS Collaboration, \emph{Probing the quantum interference between singly and doubly resonant top-quark production in pp collisions at $\sqrt{s}$ = 13 \TeV \ with the ATLAS detector}, Phys. Rev. Lett. {\bf 121} (2018) 152002 [\href{https://arxiv.org/abs/1806.04667}{\tt arXiv:1806.04667v2}].

\bibitem{ATLAS_tt_Wt_QI_prediction}
T. Je\v{z}o {\it et al.}, \emph{An NLO+PS generator for $t\bar{t}$ and $\Wboson t$ production and decay including non-resonant and interference effects}, Eur. Phys. J. C {\bf 76} (2016) 691 [\href{https://arxiv.org/abs/1607.04538}{\tt arXiv:1607.04538}].

\bibitem{CMS_tt_dilep_new_phys}
CMS Collaboration, \emph{Search for new physics in top quark production in dilepton final states in proton-proton collisions at $\sqrt{s}$ = 13 \TeV}, Submitted to Eur. Phys. J. C (March 2019) [\href{http://cms-results.web.cern.ch/cms-results/public-results/publications/TOP-17-020/index.html}{\tt CMS-TOP-17-020}].

\bibitem{CMS_tZq}
CMS Collaboration, \emph{Measurement of the associated production of a single top quark and a $\Zboson$ boson in pp collisions at $\sqrt{s}$ = 13 \TeV}, Phys. Rev. Lett. B {\bf 779} (2018) 358 [\href{https://arxiv.org/abs/1712.02825}{\tt arXiv:1712.02825v2}].

\bibitem{ATLAS_tZq}
ATLAS Collaboration, \emph{Measurement of the production cross-section of a single top quark in association with a $\Zboson$ boson in proton-proton collisions at $\sqrt{s}$ = 13 \TeV \ with the ATLAS detector}, Phys. Lett. B {\bf 780} (2018) 557 [\href{https://arxiv.org/abs/1710.03659}{\tt arXiv:1710.03659v2}].

\bibitem{CMS_tZq_obs}
CMS Collaboration, \emph{Observation of single top quark production in association with a $\Zboson$ boson in proton-proton collisions at $\sqrt{s}$ = 13 \TeV}, Phys. Rev. Lett. {\bf 122} (2019) 132003 [\href{https://arxiv.org/abs/1812.05900}{\tt arXiv:1812.05900v2}].

\end{thebibliography}
\end{document}